\documentclass[fleqn,usenatbib]{mnras}

\usepackage{newtxtext,newtxmath}

\usepackage[T1]{fontenc}
\usepackage{ae,aecompl}

\usepackage{graphicx}	
\usepackage{amsmath}	
\usepackage{subfiles}   
\usepackage{stfloats}

\title[Tracking the orbit of unresolved subhalos ]{Tracking the orbit of unresolved subhalos for semi-analytic models}

\author[Delfino et al.]{Facundo M. Delfino,$^{1,2}$\thanks{E-mail: fdelfino@fcaglp.unlp.edu.ar}
Claudia G. Sc\'occola,$^{1,2}$ 
Sof\'ia A. Cora$^{1,2,3}$,
\newauthor
Cristian A. Vega-Mart\'inez$^{4,5}$ and Ignacio D. Gargiulo$^{2,3}$
\\
$^{1}$ Facultad de Ciencias Astron\'omicas y Geof\'isicas, Universidad Nacional de La Plata, Observatorio Astron\'omico, Paseo del Bosque, \\
B1900FWA La Plata, Argentina \\
$^{2}$ Consejo Nacional de Investigaciones Cient\'ificas y T\'ecnicas (CONICET), Rivadavia 1917, Buenos Aires, Argentina \\
$^{3}$ Instituto de Astrof\'isica de La Plata (CCT La Plata, CONICET, UNLP), Observatorio Astron\'omico, Paseo del Bosque
B1900FWA, La Plata, Argentina \\
$^{4}$ Instituto de Investigaci\'on Multidisciplinar en Ciencia y Tecnolog\'ia, Universidad de La Serena, Ra\'ul Bitr\'an 1305, La Serena, Chile \\
$^{5}$ Departamento de Astronom\'ia, Universidad de La Serena, Av. Juan Cisternas 1200 Norte, La Serena, Chile
}

\date{Accepted XXX. Received YYY; in original form ZZZ}

\pubyear{2020}

\begin{document}
\label{firstpage}
\pagerange{\pageref{firstpage}--\pageref{lastpage}}
\maketitle

\begin{abstract}
We present a model to track the orbital evolution of “unresolved subhaloes” (USHs) in cosmological simulations. USHs are subhaloes that are no longer distinguished by halo finders as self-bound overdensities within their larger host system due to limited mass resolution. These subhaloes would host “orphan galaxies” in semi-analytic models of galaxy formation and evolution (SAMs).
Predicting the evolution of the phase-space components of USHs is crucial for the adequate modelling of environmental processes, interactions and mergers implemented in SAMs that affect the baryonic properties of orphan satellites.
Our model takes into account dynamical friction drag, mass loss by tidal stripping and merger with the host halo, involving three free parameters. To calibrate this model, we consider two DM-only simulations of different mass resolution (MultiDark simulations). The simulation with higher-mass resolution ({\sc smdpl}; $ m_{\rm DM} = 9.6 \times 10^7 ~ h^{-1}\,\mathrm{M_{\odot}}$) provides information about subhaloes that are not resolved in the lower-mass resolution one ({\sc mdpl2}; $ m_{\rm DM} = 1.5 \times 10^9 ~ h^{-1}\,\mathrm{M_{\odot}}$); the orbit of those USHs is tracked by our model. We use as constraining functions the subhalo mass function (SHMF) and the two-point correlation function (2PCF) obtained from {\sc smdpl}, 
being the latter a novel aspect of our approach.
While the SHMF fails to put tight constraints on the efficiency of dynamical friction and the merger condition, the addition of clustering information helps to specify the parameters of the model related to the spatial distribution of subhaloes. Our model allows to achieve good convergence between the results of simulations of different mass resolution, with a precision better than 10 per cent for both SHMF and 2PCF.
\end{abstract}

\begin{keywords}
galaxies:haloes -- galaxies:formation -- galaxies:evolution --  methods: numerical
\end{keywords}


\section{Introduction}

\label{intro}

\noindent The observed Universe is successfully described by the $\Lambda$CDM model. According to this concordance model, at early times, the Universe underwent a period of exponential expansion, called Inflation, in which the primordial perturbations in the metric were settled. These metric fluctuations produced perturbations in the matter density field that are characterised by the matter power spectrum. The large scale structure (LSS) we see today is the result of the gravitational growth of these tiny matter perturbations. It is currently accepted that structure formation proceeds in a hierarchical way, with small structures being the first ones to collapse and reach a state close to virial equilibrium. Larger structures, like massive dark matter (DM) haloes, form later by mergers of pre-existing virialised haloes, and by accretion of diffuse dark matter \citep{Frenk_White_DarkMatter_2012}.

Galaxies are born and evolve within the DM haloes \citep{White_Rees_GalaxyFormation_1978}. They are highly non-linear objects that are the result of a complex formation mechanism, which involves several astrophysical processes and spans a wide range of spatial scales~\citep[for a review of the theory of galaxy formation and evolution see, e.g.][]{Benson_GalFormTheory_2010, Silk_GalFormReview_2012,SomervilleDave_2015, Wechsler_Tinker_DMHalosReview_2018}. As a result of the non-linear nature of these processes, it is not feasible to treat them using analytical methods and hence the use of numerical simulations is required.

In the literature, there are various methods for generating simulated galaxy populations. Halo Occupation Distribution (HOD, e.g \citealt{Peacock_HOD_bias_2000,Berlind_HOD_2002,Berlind_HOD_physics_2003}) and Sub-Halo Abundance Matching (SHAM, e.g.  \citealt{Kravtsov_SHAM_2004, Vale_Ostriker_SHAM_luminosity_2004, Conroy_SHAM_luminosity_2006, Behroozi_SHAM_SM-HM_2010, Moster_SHAM_SM-HM_2010, Trujillo-Gomez_HaloAbundanceMatching_2011, Reddick_SHAM_ConnectionGalaxiesDM_2013}) are methods that specify the connection between DM halos (or subhaloes) and galaxies in a purely statistical way. Since these methods do not attempt to model the physics of galaxy formation, they are able to obtain large-volume galaxy catalogues at a low computational cost. However, it remains difficult to move from this statistical prescription to a physical understanding of the galaxy formation process itself.

On the other hand, for large scales, cosmological hydrodynamical simulations are capable of evolving the initial DM and baryon content in a direct way, while the complex sub-grid processes (such as star formation or feedback) are included using physically motivated prescriptions. At present, the state of the art of this class of simulations includes the HORIZON-AGN \citep{Dubois_HORIZON-AGN_2014} simulation, the Magneticum Pathfinder simulation \citep{Hirschmann_Magneticum_2014}, the BAHAMAS project \citep{McCarthy_BAHAMAS_2017}, the Virgo Consortium's EAGLE project \citep{Schaye_EAGLEproject_2015,Crain_EAGLE_calibration_2015} and The Next Generation Illustris project (IllustrisTNG, \citealt{Springel_IllustrisTNG_clustering_2018, Nelson_IllustrisTNG_bimodality_2018, Pillepich_IllustrisTNG_galaxies_2018, Naiman_IllustrisTNG_chemical_evolution_2018, Marinacci_IllustrisTNG_magnetic_fields_2018}). 
Although these cosmological simulations have successfully  reproduced important features observed in galaxy studies, such as the morphology of galaxies at low redshift \citep{Dubois_HorizonMorphology_2016}, the galaxy color bimodality distribution at low redshift \citep{Nelson_IllustrisTNG_bimodality_2018}, and abundance relations of individual chemical elements \citep{Naiman_IllustrisTNG_chemical_evolution_2018}, they still present limitations. Firstly, given the large dynamical range required in both mass and spatial scale, these simulations are usually extremely demanding on computing power. Secondly, their volumes are relatively small compared to current and upcoming galaxy surveys such as eBOSS \citep{Dawson_eBOSS_2015, Dawsom_eBOSS_paper_2016, Alam_eBOSS_completed_2020}, LSST \citep{Abell_LSST_Book_2009, Abate_LSST_collaboration_2012, LSST_collaboration_2018}, DESI \citep{Levi_DESI_2013,DESI_collaboration_2016} or Euclid \citep{Laureijs_Euclid_RBook_2011, Amendola_Euclid_2013, Blanchard_Euclid_collaboration_2019}, making it difficult to examine the distribution of galaxies on the baryon acoustic oscillations scale\footnote{\cite{Colombi_LSSVolumEffects_1994} shows that the two-point correlation function begins to deviate from theory when $ r \gtrapprox 0.2 L_{\mathrm{box}} $ (here $ r $ is the correlation scale and $ L_{\mathrm{box}} $ is the simulation box size).}.

Semi-analytic models (SAMs, e.g. \citealt{Springel_SAM_2001, Croton_SAGE_2006, Croton_2016, Cora_SAG_2006, Cora_SAG_2018, Somerville_2008, Benson_GALACTICUS_2012, Henriques_LGalaxies_2013, Gonzalez-Perez_GALFORM_2014}) are a good alternative to overcome the disadvantages of the previous methods. The cornerstone of SAMs are the merger tree, i.e. a statistical representation of the growth of DM haloes. Merger trees can be constructed either analytically, using the extended Press-Schechter theory (e.g \citealt{Bond_ExcursionSetTheory_1991,Somerville_MergerTrees_1999}), or be extracted from cosmological N-body simulations (e.g \citealt{Roukema_MergerTrees_1997,Tweed_N-bodyMegerTrees_2009}). The haloes (and subhaloes) obtained from the merger tree, are then populated with simulated galaxies, where the complex physical processes related to their formation and evolution are treated using semi-analytic approximations that involve free parameters. Due to their flexibility and relatively low computational costs, SAMs represent an ideal tool to build simulated galaxy populations in cosmological volumes. In addition, they enable to test the physical processes that determine the evolution of galaxies. For a review of semi-analytic methods see e.g. \cite{Baugh_SAMsReview_2006}.

Since cosmological {\em N}-body simulations can handle a large dynamic range in mass and spatial resolution at a relatively low computational cost, mock catalogues generated using SAMs based on numerical merger trees have the advantage of allowing a straightforward comparison to observational data and a fast exploration of the parameter space. Despite these advantages, a difficulty related to this approach is that DM substructures are fragile systems that may be rapidly destroyed. 
According to \cite{Han_SubhaloModel_2016}, about 45 per cent of the subhaloes in the Aquarius simulations are disrupted regardless their infall mass. A similar result has been reported for the Bolshoi simulations \citep{Klypin_Bolshoi_2011}, where about 60 per cent of the subhaloes with masses greater than 10 per cent of the host mass survive for less than one orbit \citep{Jiang_DMSubstructureStatistic_2017}. 
Besides, \cite{vdBosch_Ogiya_2018} find that most disruptions of subhaloes identified in cosmological simulations are numerical in origin.
Furthermore, halo finder algorithms have difficulties to detect over-densities located in dense regions of their host system \citep{Muldrew_SubhaloDetectAccuracy_2011}.
As a consequence of these numerical issues, most SAMs follow the evolution of satellite galaxies that have lost their DM subhalo before their eventual merger with their host. These satellites are called ``orphan galaxies''.

The necessity of including orphan galaxies has been already pointed out in previous studies. \cite{Kitzbichler_TPCFGalMergerRate_2008} used a SAM to show that, without the addition of orphan satellites, the clustering at low scales is much lower than that observed in actual galaxy surveys. 
\cite{Guo_SHAM_2PCF_2014} show that the inclusion of orphan satellites improves the numerical converge of the properties of the galaxy population in SHAM catalogues built from cosmological {\em N}-body simulations of different mass resolutions.  
Using different SAMs and HOD models, \cite{Pujol_SAMs_2PCF_Comparison_2017} studied the impact that different treatments of the population of orphan satellites have on the clustering signal. In general, it is found that models that do not include orphans present a lower clustering at low scales. 
It is worth noticing that those  models that include orphan satellites treat them in a rather simple way, considering the 
orbital decay time-scale due to dynamical friction
(\citealt{Chandrasekhar_DynamicalFriction_1943A, Boylan-Kolchin_MergingTimescale_2008} ) to estimate the moment in which they merge with the central galaxy of their host structure. Such approach prevents the model from an adequate and physically motivated prediction of their positions and velocities; while in some cases they are estimated assuming a circular orbit with a  decaying radial distance determined from the dynamical friction, with position and velocity components randomly generated (\citealt{Cora_SAG_2006, Gargiulo_SAG_downsizing_2015}), in other SAMs they are simply traced by those of the most bound particles of substructures at the last time they were identified (e.g. \citealt{DeLuciaBlaizot_2007, Benson_GALACTICUS_2012, Gonzalez-Perez_GALFORM_2014}).

A proper treatment of the positions and velocities of satellite galaxies is crucial. The phase space information is not only relevant for clustering analysis but it is also an essential ingredient for the proper estimation of interactions and mergers, and the effects of environmental processes, such as ram pressure stripping  \citep{Gunn_Gott_SphericalCollapse_1972} and tidal stripping  \citep{Merritt_1983}.
Both ram pressure and tidal stripping produce mass loss (only gas loss in the case of ram pressure stripping) as the satellite galaxy moves within a high density environment, affecting their properties (gas content, star formation rate, size, colour). 
The position and velocity of galaxies generated by a SAM is determined by the position and velocity of the (sub)halo they reside in. 
A subhalo orbiting within its host is subject to tidal forces that cause 
it to lose mass by 
tidal stripping.
This mechanism depends strongly on the circularization of the orbit, where most of the mass is lost when the halo passes through the pericentre. Also, tidal shocks at the pericentre of the orbit increase the kinetic energy and the 
subhalo 
expands (tidal heating, e.g. \citealt{Spitzer_TidalHeating_1958,Gnedin_TidalHeating_1999,Banik_VdBosch_TidalHeating_2021}) making it more susceptible to tidal stripping  \citep{Zentner_Bullock_HaloSubstructure_2003, Gan_SubhaloDynamicalEvolution_2010, Pullen_SubhaloNonlinearEvolution_2014}.
The successive decrements of the subhalo mass impact directly on the deceleration in the direction of its motion caused by dynamical friction and, consequently, on its orbital evolution.
When the subhalo is no longer detected, these processes must be taken into account to model the orbital evolution of the orphan galaxy generated by a SAM. In other words, an adequate treatment of the orphan galaxies in a SAM requires tracking the orbital evolution of their corresponding vanished DM subhaloes.

Throughout this paper we refer to these ``dissapeared'' subhaloes which are hosts of orphan galaxies as \textit{unresolved subhaloes}.  
The purpose of this work is to present an improved treatment for the orbital evolution of unresolved subhalos that sink into their host's potential, to be used in SAMs. This model includes the effects of tidal stripping, dynamical friction and a criterion to determine whether an unresolved subhalo is merged with its host; each of these processes involves a free parameter. 
The modelling of these physical processes follows an approach similar to those adopted in previous studies \citep[]{Taylor_Babul_SatGalDynamics_2001,Zentner_SubhaloModel_2005,Peniarrubia_Benson_SubhaloDynamicalEvolution_2005,Gan_SubhaloDynamicalEvolution_2010,Pullen_SubhaloNonlinearEvolution_2014}.
However, these previous works focus on Milky Way-sized host haloes to calibrate the free parameters involved in the modelling. A model for the orbital evolution of unresolved subhaloes that is designed to run as a pre-processing step before applying a SAM on large cosmological volumes needs to be calibrated taking into account orbits of satellite haloes (subhaloes) within host DM haloes that cover a larger dynamical range. Such calibration can be carried out by extracting results from cosmological simulations or accretion histories obtained via extended Press Schechter theory, such as individual orbits of subhaloes or collective statistical quantities (e.g. the halo mass function).
Recently, \cite{Yang_SubHaloMCMC_2020} presented a calibration method for the parameters of their subhalo orbital evolution model that uses as contraints the mass function and circular velocities of subhaloes extracted from cosmological simulations.
However, the lack of massive subhalos in their sample  prevents the method from putting tight constraints on the dynamical friction model, which affects mainly massive subhaloes. Moreover, the statistics used do not take into account the spatial distribution of subhaloes that would help to put constraints at small scales.

We propose a different calibration procedure. In order to tune the free parameters of our model for the orbital evolution of unresolved subhalos, we consider two DM-only cosmological simulations with same settings but different box size and mass resolution. We choose a
subvolume 
of the lower-resolution simulation and track the positions and velocities of those subhaloes that become unresolved by applying the orbital evolution model. 
Its free parameters are tuned by considering as constraints the subhalo mass function and the two-point correlation function given by the higher-resolution simulation, since subhaloes that are not present in the lower-resolution simulation would be identified in the higher-resolution one.
We assume that any difference between the constraining functions obtained from the two DM-only simulations will be produced by the orbits of unresolved subhaloes, and we tune the parameters of the orbital evolution model by minimizing such differences. Since the correlation function contains information on the spatial distribution of subhaloes, its inclusion as constraint could help to better define the dynamical friction model and the merger condition.

This paper is organised as follows: in Section~\ref{model}, we introduce the orbital evolution model for unresolved suhaloes. Section~\ref{methodology} describes the statistical tools used to analyse galaxy clustering and gives details about the simulations used in this work, the MultiDark simulations \textsc{mdpl2} and \textsc{smdpl}. In Section~\ref{results}, we  describe the method used to calibrate the free parameters of the model. In Section~\ref{results:ValidationProcedure}, we present an analysis of individual orbits to assess the validity of the calibration procedure. We discuss our results in Section~\ref{discussion}, and present our conclusions in Section~\ref{conclusions}.

\section{Orbital evolution model for unresolved subhaloes}
\label{model}

A common practice followed by SAMs to generate a galaxy population is to take as input the properties of DM haloes and their merger trees extracted from cosmological {\em N}-body simulations \citep[see e.g.][]{Roukema_MergerTrees_1997,Springel_SAM_2001,Croton_SAGE_2006,Cora_SAG_2018}. Based on this information, the model assigns a central galaxy to each new halo that appears in the DM simulation and then follows the evolution of its properties. 
When a smaller halo falls into a larger one, so that it can be identified as a substructure, the corresponding central galaxy becomes a satellite one.
For satellite galaxies with well defined DM subhaloes, effects such as dynamical friction or tidal stripping are given in a self-consistent way by the base {\em N-}body simulation. 
Instead, orphan satellite galaxies are associated to unresolved subhaloes (hereafter USHs), and their positions and velocities cannot be followed accurately. 
When a subhalo is no longer detected in the simulation, different assumptions and modelling  have to be assumed to continue tracking its orbit. As we have discussed in Section~\ref{intro}, those SAMs that include orphan galaxies have taken different approaches to deal with this issue.

The simplest model of a subhalo moving within its host halo can be approximated by a point mass without internal structure orbiting in a static potential. This model does not take into account the internal structure of the subhalo or the interactions with the material that forms the host halo, which are relevant for its orbital evolution. 
For example, the interaction between the subhalo and the matter of its host gives rise to a dynamical friction force that causes the evolution of the orbits to deviate from that of the simplest model. In addition, during its evolution, a 
subhalo 
may experience mass loss due to tidal stripping or gravitational shocks. Therefore, a more physically-motivated description of the orbital evolution of USHs calls for a model that takes into account the aforementioned processes.

We present a model to track the orbits of USHs that can be used in a pre-processing step within a SAM pipeline, that is, before applying a semi-analytic model of galaxy formation to the underlying cosmological DM-only simulation; this procedure has been followed with the semi-analytic model SAG (\citealt[][ see their section 3.2.]{Cora_SAG_2018}).
We consider each USH as an object moving in a smooth spherical potential generated by its host halo. The initial conditions to integrate its orbit are the position, velocity, mass and radius of the subhalo at the instant of last identification, as given by the cosmological {\em N}-body simulation.
In order to take into account the effects of the host halo over the dynamics of 
an USH, 
at each instant, we compute the effect of dynamical friction using Chandrasekhar's formula and we also take into account mass loss by considering a tidal stripping model. If the mass of an 
USH falls below a certain resolution limit then we consider it as disrupted; if the distance of an USH 
to the centre of its host becomes less than a fraction of the viral radius of the host or 
its specific angular momentum is less than the allowed minimum, 
then the USH is considered to be merged with its host. Below, we describe these processes in more detail.

\subsection{Dynamical friction (DF)}\label{model:DynamicalFriction}

When  
a subhalo 
of total mass 
$M_{\mathrm{sub}}$ 
moves through a large collisionless system composed of particles of mass 
$ m << M_{\mathrm{sub}} $, 
it perturbs the particle field creating an over-dense region behind it. This ``wake'' pulls the subhalo in the opposite direction causing a drag force called dynamical friction (DF, hereafter).
Therefore, 
when tracing the dynamics of an USH of mass $M_{\mathrm{ush}}$ 
orbiting within a massive halo, we can separate the force experienced by it 
into two contributions: one due to the potential of the host halo, and a higher-order correction due to the background particles (the dynamical friction term). The first part is given by $f_{i} = - \partial \Phi / \partial x_{i}$, that relates the force acting on a particle at a given position with the potential of the main system $ \Phi $ at that position. 
Here, we assume that $ \Phi $ corresponds to a mass density radial distribution that follows a Navarro-Frenk-White profile \citep[NFW,][]{Navaro_NFW_1997}. For more details on the NFW density profile, see Appendix \ref{appendixA} and \cite{Lokas_Mamon_NFWprofile_2001}.

The DF force is given by the Chandrasekhar formula 
\citep{Chandrasekhar_DynamicalFriction_1943A, Binney_Tremaine_GalDynamics}, i.e.

\begin{equation}
\begin{split}
    \textbf{F}_{\rm ush}^{\rm df} = &- \frac{4 \pi G^2 M_{\mathrm{ush}}^2 \, \rho_{\rm host}(r_{\rm ush}) \ln{(\Lambda}) }{V_{\rm ush}^{2}} \times \\ &\Bigg[\mathrm{erf}(X) 
    -\frac{2 X}{\sqrt{\pi}} \exp(-X^{2}) \Bigg] \frac{\textbf{V}_{\rm ush}}{V_{\rm ush}}~,
    \label{model:Eq:Chandrasekhar1943}    
\end{split}
\end{equation}

\noindent where 
$ r_{\rm ush} $ 
is the position of the USH relative to its host halo, 
$\textbf{V}_{\rm ush}$ 
is the velocity of the USH, 
$V_{\rm ush} = | \textbf{V}_{\rm ush} |$, 
$ X = V_{\rm ush} / (\sqrt{2} \sigma) $ 
with $ \sigma $ the velocity dispersion of dark matter particles in the host halo, 
$ \rho_{\mathrm{host}} $ 
represents the mass density distribution of the host halo, $ \ln{(\Lambda)} $ is the Coulomb logarithm and $\mathrm{erf}$ is the Gauss error function.
Assuming a NFW profile for the density distribution 
$ \rho_{\mathrm{host}} $ 
and an isotropic velocity distribution, the velocity dispersion $ \sigma $ is given by

\begin{equation}
    \frac{\sigma^{2}(x)}{V_{\rm vir}^{2}} = c \, g(c) \, x \, (1 + x)^{2} \int_{x}^{\infty} \left[ \frac{1}{g(x') \, {x'}^{3} \, (1 + x')^{2} } \right] dx'~,
    \label{model:Eq:NFWsigma}
\end{equation}

\noindent where  $ x = c \, s $, $ c $ is the concentration parameter
of the host, 
$ 1/g(x) = \ln(1 + x) - x/(1 + x) $, and $ s $ is the distance from the center of the host halo normalised by its virial radius.  
For simplicity, in this work we use the following approximation for $ \sigma $, which is accurate to 1 per cent for $x$ in the range $ 0.01 - 100$ \citep{Zentner_Bullock_HaloSubstructure_2003}

\begin{equation}
    \sigma(x) \simeq V_{\mathrm{max}} \frac{1.4393 ~ x^{0.354}}{ 1 + 1.1756 ~ x^{0.725} }~,
    \label{model:Eq:NFWsigmaAprox}
\end{equation}

\noindent where $V_{\mathrm{max}}$ is the maximum circular velocity
of the host, 
related to the virial velocity
via 
$ V_{\mathrm{max}} \simeq V_{\mathrm{vir}} \sqrt{0.216~c~g(c)} $.

The argument of the Coulomb logarithm can be expressed as $ \Lambda = b_{\mathrm{max}} / b_{\mathrm{min}} $ where $ b_{\mathrm{max}} $ and $ b_{\mathrm{min}} $ are the maximum and the minimum impact parameters for gravitational encounters between 
a subhalo 
and the background objects \citep{Binney_Tremaine_GalDynamics}. Typically, $ b_{\mathrm{min}} $ corresponds to a close encounter, then 
$ b_{\mathrm{min}} \simeq G M_{\rm sub} / V^2 $ 
where  $ V $ is a velocity typical of the encounter, such as the rms velocity of the background particles. The choice of the value for $ b_{\mathrm{max}} $ is more ambiguous, and for a finite system is taken to be the characteristic scale of the system.

It should be noted that the derivation of Chandrasekhar's formula assumes a massive particle moving in a homogeneous medium composed by an infinite number of low-mass particles with a Maxwellian velocity distribution. However, in the literature, several works show that this equation is applicable to more general contexts, where these hypotheses are not satisfied, if the Coulomb logarithm is chosen appropriately \citep{Weinberg_SatGalOrbitalDecay_1986, Cora_OrbitalDecayDF_1997, Velazquez_White_DynFriction_1999}.

There has been much debate in the literature about the appropriate choice of Coulomb logarithm. For example, \cite{Springel_SAM_2001} uses an approach given by 
$ \ln (\Lambda) \simeq \ln(1 + M_{\mathrm{cen}}/M_{\mathrm{sub}}) $ 
where 
$ M_{\mathrm{cen}} $ and 
$ M_{\mathrm{sub}} $ 
are the masses of the central halo 
and its 
subhalo, 
respectively. On the other hand, some authors use other definitions that allow them to reproduce results from {\em N}-body simulations \citep{Hashimoto_OrbitCircularization_2003, Zentner_Bullock_HaloSubstructure_2003, Petts_DynFrictionCore_2015, Petts_DynFrictionCored_2016,  Ogiya_Burkert_DynFriction_2016}. In particular, \cite{Hashimoto_OrbitCircularization_2003} propose a variable Coulomb logarithm. This choice avoids the strong circularization effect that is observed when comparing these models with the results obtained from {\em N}-body simulations. Following this, we use the expression

\begin{equation}
    \ln (\Lambda) = 
        \begin{cases}
            \ln(r_{\mathrm{ush}}/b R_{\mathrm{ush}}) &r_{\mathrm{ush}} > b R_{\mathrm{ush}} \\
            0  &r_{\mathrm{ush}} \leq b R_{\mathrm{ush}}\\
        \end{cases}~,
    \label{model:Eq:LnCHashimoto2003}
\end{equation}

\noindent where 
$ r_{\rm ush} $ 
is the distance of the USH to the centre of its host halo, 
$ R_{\rm ush} $ 
is the instantaneous virial radius of the USH, and $ b $ is a free parameter. 
$R_{\rm ush}$ 
is calculated by considering that it encloses the mass of the USH (which is estimated at certain moments during the integration process and becomes progressively reduced as the result of tidal stripping), and requiring that the corresponding uniform density is  200 times the critical density of the Universe at that epoch.
Note that in \citet{Hashimoto_OrbitCircularization_2003}, the Coulomb logarithm is given by 
$ \ln (\Lambda) = \ln(r/1.4 \epsilon) $ 
for 
$r > 1.4 \epsilon$, 
where 
$ \epsilon $ 
is the softening length corresponding to a Plummer sphere. Here, we assume a NFW profile for the 
USH, 
thus we introduce 
its 
virial radius 
and leave $ b $ as a free parameter to be adjusted.

\subsection{Tidal stripping (TS)}\label{model:TidalStripping}

As mentioned above,
a subhalo 
orbiting within its host system is subjected to tidal forces. When tidal forces are greater than the gravitational force of the 
subhalo 
itself, 
part of 
the material becomes unbound and the 
subhalo 
loses mass. The DF force is proportional to 
the mass squared 
(see equation~\ref{model:Eq:Chandrasekhar1943}), and hence the magnitude of the deceleration experienced by 
a subhalo 
is proportional to 
its mass. 
As a result, mass loss can have a major impact on the orbital evolution of the 
subhalo. 
For this reason, it is necessary to estimate the amount of mass lost by tidal stripping (TS, hereafter) 
of an USH.

We estimate the tidal radius as the distance at which the self-gravity force and the tidal forces cancel out; material outside this distance becomes unbound and could be stripped out from 
an 
USH. The tidal radius is given by

\begin{equation}
    r_{\mathrm{t}} = \left( \frac{G M_{\rm ush}}{\Omega_{\rm ush}^{2} - d^2 \Phi / d r^2} \right)^{1/3}~,
    \label{model:Eq:TidalRadiusKing}
\end{equation}

\noindent where 
$ M_{\rm ush} $ 
is the mass of the USH, 
$ \Omega_{\rm ush} $ 
is its angular velocity and $ \Phi $ characterises the potential of the host system \citep{King_TidalRadius_1962,Taylor_Babul_SatGalDynamics_2001,Zentner_Bullock_HaloSubstructure_2003}. This equation is derived under the assumption that the
USH 
moves in a circular orbit and the potential of the main system is spherically symmetric. But, even under these restricted assumptions, the tidal limit cannot be represented as a spherical surface because some particles 
within $ r_{\mathrm{t}} $ are unbound while other particles outside $ r_{\mathrm{t}} $ may remain bound to the 
USH 
\citep{Binney_Tremaine_GalDynamics}.

In general, 
USHs 
do not move in circular orbits and the potential of the host system is not spherically symmetric. 
As an approximation, we can still apply equation~\ref{model:Eq:TidalRadiusKing} to eccentric orbits, in which case we estimate an instantaneous tidal radius by using the corresponding instantaneous values, i.e. 
$ \Omega_{\rm ush} = | \textbf{V}_{\rm ush} \times \textbf{r}_{\rm ush} | / r_{\rm ush}^2 $, 
where 
$ \textbf{V}_{\rm ush} $ 
and 
$ \textbf{r}_{\rm ush} $ 
are the instantaneous position and velocity of 
the 
USH.
The final expression of the tidal radius obtained by assuming a NFW mass density profile for both the
USH 
and the host halo is given in the Appendix~\ref{appendixA} (equation~\ref{appendixA:Eq:NFW_rt}).

Another aspect that remains unclear is the rate at which the material located outside the instantaneous tidal radius $ r_{\mathrm{t}} $ is going to be removed. Following \cite{Zentner_SubhaloModel_2005}, we absorb all these complicated details in a free parameter to be adjusted by external constraints. Then, the rate of mass loss of an USH by TS is given by

\begin{equation}
    \frac{d M_{\rm ush}}{d t} = - \alpha \frac{M_{\rm ush}(>r_{\mathrm{t}})}{T_{\mathrm{orb}}}~.
    \label{model:Eq:TSrate}
\end{equation}

\noindent Here 
$M_{\rm ush}(> r_{\rm t})$ 
is the mass of the USH outside the tidal radius $ r_{\rm t} $,
$ T_{\rm orb} = 2 \pi / \Omega_{\rm ush} $, 
with 
$ \Omega_{\rm ush} $ 
the instantaneous angular velocity of the USH, and $ \alpha $ is treated as a free parameter. The value of the parameter $ \alpha $ differs from author to author. For example, \cite{Taylor_Babul_SatGalDynamics_2001} and \cite{Zentner_Bullock_HaloSubstructure_2003} choose a value $ \alpha = 1 $; on the other hand, \cite{Peniarrubia_Benson_SubhaloDynamicalEvolution_2005} assume an instantaneous stripping, which effectively implies $ \alpha \rightarrow \infty $. Some authors (e.g. \citealt{Zentner_SubhaloModel_2005}, \citealt{Pullen_SubhaloNonlinearEvolution_2014}) vary the value of $ \alpha $ in order to reproduce the halo mass function of numerical simulations.

\subsection{Merger criterion}\label{model:MergerCriterion}

According to hierarchical structure formation models, mergers play a critical role in the formation and evolution of galaxies. 
When two (sub)haloes merge to give place to a larger structure, the smaller one is no longer detected, becoming 
an 
USH.   
In the scheme implemented by SAMs, the central galaxy of the larger progenitor of the remnant (sub)halo becomes its central galaxy, while the central galaxy of the smaller progenitor becomes an orphan satellite. Those SAMs that include the modelling of orphan satellites need to estimate their merger time-scale with the corresponding central galaxy. 
It has been common practice to estimate a dynamical friction time-scale $t_{\rm DF}$ (e.g. \citealt{Chandrasekhar_DynamicalFriction_1943A, Boylan-Kolchin_MergingTimescale_2008, Jiang_MergerTimescale_2008}) at the moment in which the smaller progenitor (sub)halo merges with the larger one, that is, when it becomes an USH, and to consider the subsequent evolution of the galaxy contained in that dissapeared (sub)halo (orphan satellite) until the time $t_{\rm DF}$ has elapsed. When that condition is fulfilled, the merger between the orphan satellite and its central galaxy takes place. Note that the dynamical friction time-scale can be reset if the (sub)halo in which the orphan satellite orbits falls into a larger one. The drawback of this kind of implementation is that it does not provide any information about the position and velocity of the orphan satellite, and further assumptions have to be made (see Section~\ref{intro}). 
Thus, a model of the orbital evolution of USHs is an useful tool to be applied on DM halo merger trees as a pre-processing step, before modelling the evolution of the galaxy population by SAMs \citep[e.g.,][]{Cora_SAG_2018, Cora_2019}.
A criterion to determine whether 
an 
USH is merged with the central part of its host is another important aspect to take into account in such models.

In our orbital evolution model, we assume that 
an 
USH effectively merges with its host halo when it reaches the central part of the host; we do not allow the USH to pass through this region and continue moving along its orbit.
This situation occurs either when the 
specific angular momentum of the USH is reduced below 
a minimum value chosen as low as 
$0.01 \, h^{-1} \mathrm{kpc} \, \mathrm{km \, {s}^{-1}}$, 
or when the distance of the USH to the centre of its host halo,
$r_{\rm ush}$, 
is smaller than a fraction $f$ of the virial radius of the main system, $ R_{\mathrm{host}} $, i.e. when the condition

\begin{equation}
    r_{\mathrm{ush}} < f R_{\mathrm{host}}~,
    \label{model:Eq:f}
\end{equation}

\noindent is satisfied.  Here, we consider $ f $ as a free parameter of the model. It is taken as an estimation of the radius of a central galaxy that will populate the host halo via a SAM.

\subsection{Implementation of the model}\label{subsection:OrbitModelSummary}

Once the (sub)halo of a galaxy is no longer detected in the simulation, we cannot follow the evolution of its phase space coordinates and we flag it as an USH. 
From that moment on, we apply our orbital evolution model and integrate its orbit numerically, taking as initial conditions the last known values of position, velocity, mass and radius of the (sub)halo before becoming unresolved.
The integration is carried out by considering time intervals of lengths $ \delta t = \Delta T/N_{\rm steps}^{\rm orb}$, where $\Delta T$ is the spacing between snapshots given by the DM simulation, and $N_{\rm steps}^{\rm orb}$ is the number of steps involved in the orbit's integration. The latter number has been chosen as $N_{\rm steps}^{\rm orb}=5000$; this number is increased when needed to guarantee 
that energy and specific angular momentum do not increase along the evolution.
Note that $\delta t \ll \Delta T$. 
At each time interval $ \delta t$, the forces acting on an USH are computed according to equations (\ref{appendixA:Eq:NFW_acceleration}) and (\ref{appendixA:Eq:NFW_DF}), in  Appendix \ref{appendixA}, which are complemented by equations (\ref{model:Eq:NFWsigmaAprox}) and (\ref{model:Eq:LnCHashimoto2003}), and the 
positions and velocities are evolved by using a kick-drift-kick (KDK) leapfrog scheme.

In order to reduce the amount of calculations, we only update the value of the Coulomb logarithm (equation \ref{model:Eq:LnCHashimoto2003}) and apply TS at time intervals $ \delta t^{\rm out}= \Delta T/N_{\rm steps}$,
at which the results of the integration are stored. The number of steps $N_{\rm steps}$ is chosen to agree with the number of time-steps of equal size adopted by SAMs to subdivide intervals between simulation outputs, $\Delta T$, in order to integrate the differential equations that regulate the baryonic processes involved in galaxy evolution. The aim of this choice is to provide the input information that a SAM needs to follow the orbital evolution of an orphan satellite; in particular, we take $N_{\rm steps}=25$.

When applying TS, we evaluate if the tidal radius $ r_{\mathrm{t}} $ (see equation~\ref{appendixA:Eq:NFW_rt} in Appendix~\ref{appendixA}) estimated for an USH is smaller than its current radius. If that condition is satisfied then we assume that the material that is outside $r_{\mathrm{t}}$ is unbound and eligible to be removed by TS; note that we are assuming that the mass profile does not evolve and is simply truncated at $r_{\mathrm{t}}$.
According to equation ~\ref{model:Eq:TSrate}, the mass loss rate takes place within a time-scale $T_{\rm orb}$. Therefore, if $T_{\rm orb} < \delta t^{\rm out}$ then all the unbound mass 
$ M_{\rm ush}(> r_{\mathrm{t}}) $ 
is removed by TS, otherwise the stripped mass is given by 
$\alpha (\delta t^{\rm out}/T_{\rm orb}) M_{\rm ush}(> r_{\rm t }) $. Indeed, in general  $T_{\rm orb} > \delta t^{\rm out}$, thus supporting our choice of time interval for updating the effect of TS and the value of the Coulomb logarithm. 
Then, we update the mass of the USH and recalculate its corresponding virial radius (see equation \ref{appendixA:Eq:NFW_Mvir} in  Appendix \ref{appendixA}).

We continue tracking the orbital evolution of an USH until it eventually reaches the centre of its host (sub)halo, 
loses its angular momentum, or it becomes disrupted. The first case occurs when the USH satisfies the proximity condition (equation \ref{model:Eq:f}). The second one occurs when its specific angular momentum reaches the minimum value $ 0.01 \, h^{-1}\mathrm{kpc} \, \mathrm{km \, {s}^{-1}} $, which was arbitrarily chosen to avoid long integration times-scales.
The last one (disruption) takes place when the mass of the USH is reduced below a minimum value of
$ 10^{5} \, h^{-1} \mathrm{M_{\odot}}$. 
This value has been chosen to keep the compromise between two aspects. On the one hand, it is low enough to track subhaloes of very low masses reducing the effect of artificial subhalo disruption inherent to numerical simulations (e.g., \citealt{Green_2021}). On the other hand, it is high enough to prevent extremely long integration time-scales. 
In any case, the USH can be considered to be merged with its host halo. However, since we are interested in providing the condition for mergers that could follow an orphan satellite (galaxy within an USH) modelled by a SAM, 
hereafter, we will refer as mergers between the USH and its host
to the events of 
reaching the host centre (proximity criterion) or achieving a minimum value of angular momentum (criterion of angular momentum loss).
It is worth emphasising that the results of the calibration procedure applied to our model is not sensitive to the value of the minimum mass adopted for the disruption criterion as long as this minimum mass is
smaller
than the mass cut considered in our analysis (see next Section). However, this value will affect the number density of orphan satellites hosted by the surviving USHs as generated by a SAM, and their fate will be resolved together with the treatment of their baryonic components.

\section{Calibration methodology}\label{methodology}

In the previous section, we introduced a model to track the orbital evolution of USHs. This model depends on three free parameters: $ b, f,$ and $\alpha  $, introduced in equations \ref{model:Eq:LnCHashimoto2003}, \ref{model:Eq:TSrate} and \ref{model:Eq:f}, respectively. In this work, we propose to use as constraining functions the subhalo mass function (SHMF; $ \tilde{\phi} = d n / d \log M $, with $n$ the number density of subhaloes) and the two-point correlation function (2PCF; $\xi$) given by a DM simulation of higher mass resolution than the one on which the orbital evolution model is applied.

Therefore, we consider two dark matter only {\em N}-body simulations, with the same cosmological parameters but different mass and force resolutions.
It is worth noting that the halo finder is able to detect subhaloes in the high-resolution simulation that are not identified in the low-resolution one.
Hence, the results of the orbital evolution model applied to the low-resolution simulation can be compared to those obtained from the high-resolution one, which is considered as the reference simulation. Then, we vary the values of the free parameters of the model until we find a combination for which the constraining functions SHMF and 2PCF derived from the low-resolution simulation converges to those obtained fromthe high-resolution one.

Below, we describe the set of dark matter only simulations used in this work and the computation of the 2PCF.
In order to find optimal parameter values, we need to run the model several times. Since running the code over the whole boxes is a numerically expensive task, we make a parameter exploration and the corresponding convergence test over a set of boxes which are selected 
subvolumes 
of the full simulations with 2PCF and SHMF similar to those of the full boxes. The details of  this procedure is covered in the rest of this section.

\subsection{{\sc mdpl2} and {\sc smdpl} simulations}\label{methodology:Simulations}

In this subsection, we describe the two cosmological {\em N}-body simulations used to calibrate our model for the orbital evolution of USHs. 
These simulations are part of the MultiDark cosmological simulation suite\footnote{\url{https://www.cosmosim.org/}}: {\sc smdpl} and {\sc mdpl2}.
These simulations are characterized by the same Planck cosmological parameters: $ \Omega_{\rm m} = 0.307 $, $ \Omega_{\Lambda} = 0.693 $, $ \Omega_{\rm b} = 0.048 $, $ n_{\rm s} = 0.96 $ and $ H_{0} = 100~h\, \mathrm{km}\,\mathrm{s^{-1}}\, \mathrm{Mpc^{-1}} $, where $ h = 0.678 $ \citep{Ade_Plack2015_2016}. 
They follow the evolution of $ 3840^{3} $ particles within boxes of different size.
{\sc smdpl} simulation has a box with a side length of $400~h^{-1}\,\mathrm{Mpc}$ which implies a particle mass of $ 9.6 \times 10^7 ~ h^{-1}\,\mathrm{M_{\odot}} $, while {\sc mdpl2} simulation has a box size of $1~h^{-1}\,\mathrm{Gpc}$ on a side which implies 
DM particles of $ 1.5 \times 10^9~h^{-1}\,\mathrm{M_{\odot}} $.
Both simulations have been carried out with {\sc l-gadget-2} code, a version of the publicly available code {\sc gadget-2} \citep{Springel_GADGET1_2001, Springel_GADGET2_2005} whose performance has been optimised for simulating large numbers of particles. Table~\ref{methodology:Tab:SimulationParameters} shows the numerical and cosmological parameters for the simulations. For more details about this set of cosmological simulations see \citet{Klypin_MultiDark_2016}.

These simulations were analysed with the {\sc rockstar} halo finder \citep{Behroozi_ROCKSTAR_2013}, and merger trees were constructed using {\sc consistent-trees} \citep{Behroozi_ConsistentTrees_2013}. The virial mass of the DM haloes is defined as the mass enclosed by a sphere of radius $R_{\mathrm{vir}}$, so that the mean density is equal to $ \Delta = 200 $ times the critical density of the universe $ \rho_{\mathrm{c}} $, i.e. $ M_{\mathrm{vir}} = 4/3 \pi R_{\mathrm{vir}}^{3}  \Delta \rho_{\mathrm{c}} $. DM structures can exist over the background density or lie within another dark matter halo. To differentiate them, the former are referred to as main host haloes, whereas the latter are called subhaloes.

\begin{table*}
	\centering
	\begin{tabular}{ccccccccccc} 
		\hline
		simulation & box & $N_{\rm p}$ & $m_{\rm p}$ & $ \varepsilon $ & $ \Omega_{\rm m} $ & $ \Omega_{\rm b} $ & $ \Omega_{\Lambda} $ & $ \sigma_{8} $ & $ n_{\rm s} $ & $ H_{0} $ \\
		& $h^{-1}\,\mathrm{Gpc}$ &  & $h^{-1}\,\mathrm{M}_{\odot}$ & $h^{-1}\,\mathrm{kpc}$ &  &  &  &  & & $\mathrm{km}\,\mathrm{s^{-1}}\,\mathrm{Mpc^{-1}}$ \\
		\hline
		{\sc MDPL2} & 1.0 & $ 3830^{3} $ & $ 1.5 \times 10^{9} $ & $ 5.0 $ & $ 0.307 $ & $ 0.048 $ & $ 0.693 $ & $ 0.829 $ & $ 0.96 $ & $ 67.8 $  \\
		{\sc SMDPL} & 0.4 & $ 3830^{3} $ & $ 9.6 \times 10^{7} $ & $ 1.5 $ & $ 0.307 $ & $ 0.048 $ & $ 0.693 $ & $ 0.829 $ & $ 0.96 $ & $ 67.8 $ \\
		\hline
	\end{tabular}
	\caption{Numerical and cosmological parameters of the two DM-only cosmological {\em N}-body simulations used to calibrate the orbital evolution model.}
	\label{methodology:Tab:SimulationParameters}
\end{table*}

\begin{figure}
    \centering
    \includegraphics[width=1.0\linewidth]{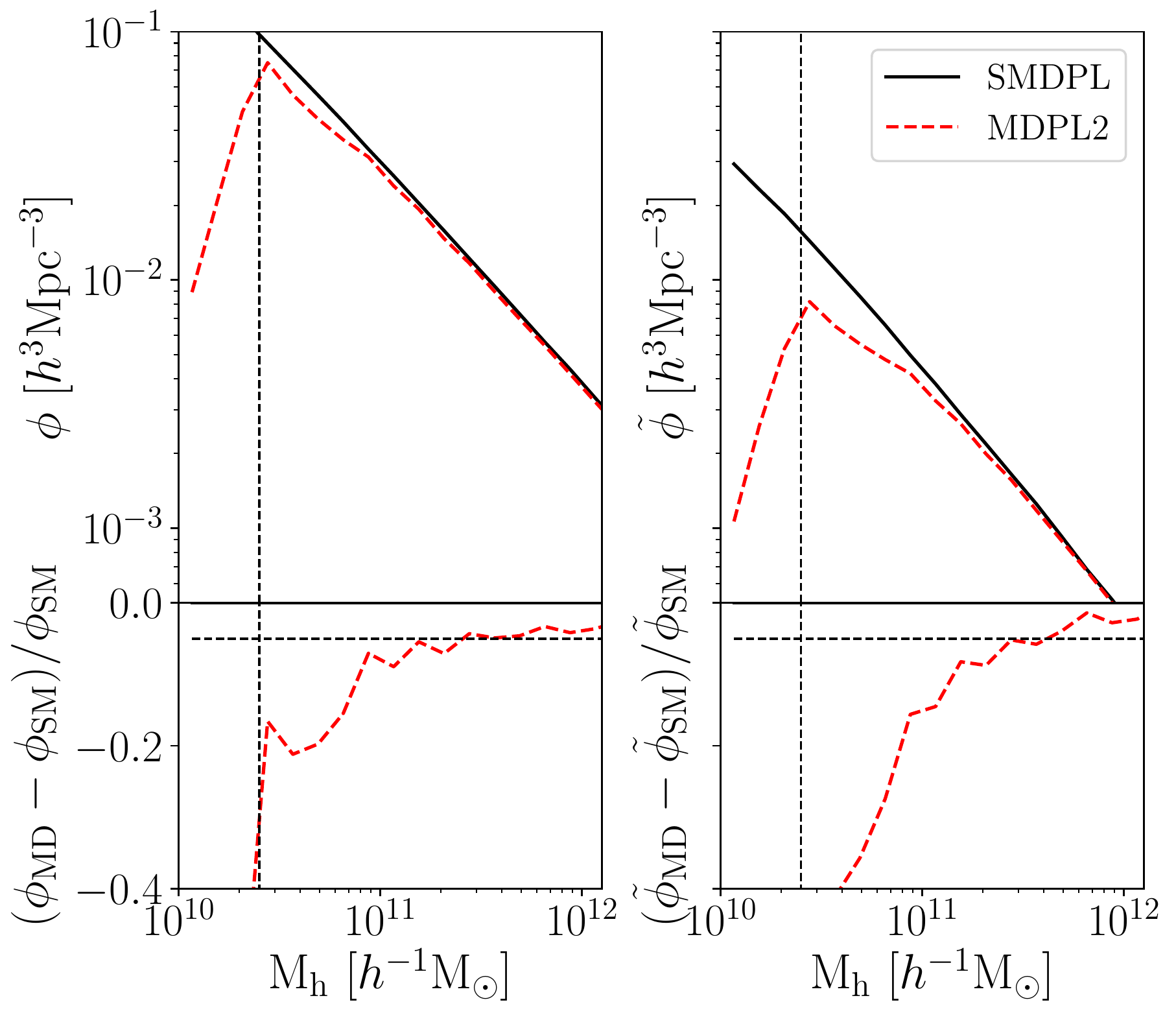}
    \caption{
    \textit{Left panels:} Halo mass function (HMF) of the simulations {\sc mdpl2}, $\phi_{\rm MD}$ (red dashed line), and {\sc smdpl},  $\phi_{\rm SM}$ (black solid line), at redshift $ z = 0 $. The lower-left panel shows the fractional difference between {\sc mdpl2} and {\sc smdpl} halo mass functions. Note that for masses lower than $10^{11}~h^{-1}\,\mathrm{M_{\odot}}$ there are approximately 20 per cent more haloes in {\sc smdpl} than in {\sc mdpl2}.
    \textit{Right panels:} Subhalo mass function (SHMF) of {\sc mdpl2}, $\tilde{\phi}_{\rm MD}$ (red dashed line), and {\sc smdpl}, $\tilde{\phi}_{\rm SM}$  (black solid line), at redshift $ z = 0 $. For masses lower than $10^{11}~h^{-1}\,\mathrm{M_{\odot}}$ there are roughly 30 per cent more subhaloes in {\sc smdpl} than in {\sc mdpl2}. In both the HMF and SHMF, the vertical dashed line denotes (sub)halo masses of $ 10^{10.4} $ $h^{-1}$ $ \mathrm{M_{\odot}} $. Note that for masses below this limit, the {\sc mdpl2} simulation presents a break which establishes the minimum mass above which we can guarantee that we have completeness in the number of haloes for both simulations. The dashed horizontal line indicates a fractional difference of $ 0.05 $.
    }
    \label{methodology:Fig:HMF_MDPL2_SMDPL}
\end{figure}

The top-left panel of Figure~\ref{methodology:Fig:HMF_MDPL2_SMDPL} shows the halo mass function (HMF; $\phi$) of the full sample of haloes (i.e., host haloes and subhaloes)
for the {\sc smdpl} ($\phi_{\mathrm{SM}}$, in solid line) and {\sc mdpl2} ($\phi_{\mathrm{MD}}$, in dashed line) simulations at redshift $ z = 0 $, while the top-right panel shows the corresponding SHMF for {\sc smdpl} ($\tilde{\phi}_{\mathrm{SM}}$, in solid line) and {\sc mdpl2} ($\tilde{\phi}_{\mathrm{MD}}$, in dashed line) for $ z = 0 $. 
As we can see from the figure, the HMF for the {\sc mdpl2} simulation presents a break at a mass of $ \sim 10^{10.4}~h^{-1} \, \mathrm{M_{\odot}} $, which establishes the minimum mass from which we can guarantee that we have completeness in the number of haloes for both simulations (vertical dashed lines in Figure~\ref{methodology:Fig:HMF_MDPL2_SMDPL}). The bottom-left panel shows the fractional difference between the HMF of the {\sc mdpl2} simulation with respect to the  one of the {\sc smdpl} simulation. In the mass range $10^{10.4} -  10^{11.0}~h^{-1}\,\mathrm{M_{\odot}}$, the fractional difference is of the order of $0.2$, hence there are $ \sim 20 $ per cent more low mass haloes in {\sc smdpl} compared with {\sc mdpl2}. On the other hand, the bottom-right panel shows the fractional differences between the SHMF functions of the {\sc mdpl2} and {\sc smdpl} simulations. In this case, for the masses below $ 10^{11.0}~h^{-1}\,\mathrm{M_{\odot}}$, we have $ \sim 30 $ per cent more subhaloes in {\sc smdpl} compared with {\sc mdpl2}. 
In both cases, for (sub)halo masses greater than $10^{11.5}~h^{-1}\,\mathrm{M_{\odot}}$, the fractional difference is always below $ \sim 0.05 $ (horizontal dashed line).

The {\sc rockstar} halo finder considers (sub)haloes formed by at least ten dark matter particles, although their properties are not robust when approaching this minimum. According to \cite{Behroozi_ROCKSTAR_2013}, halo detection is reliable for structures composed of at least twenty DM particles. 
Since completeness is important when comparing the 2PCF of different simulations, in this paper we will consider only (sub)haloes with masses greater than $10^{10.4}~h^{-1}\,\mathrm{M_{\odot}}$ for both simulations (see vertical dashed lines in Figure \ref{methodology:Fig:HMF_MDPL2_SMDPL}).

\subsection{The two-point correlation function}\label{methodology:2PCF}

Given a set of points, the probability of finding an object in an infinitesimal volume $dV$ is $dP = n dV$, where $n$ is the mean number density and $ N = n V $ is the number of objects in a finite volume $ V $. Then, the 2PCF is defined as the excess probability of finding one of them inside a small volume $ dV_{1} $ and the other in a small volume $ dV_{2} $, separated by a distance $r$ \citep{Peebles_Large_Scale_Structure, Martinez_Galaxy_Stats}, that is

\begin{equation}
    dP = n^{2} (1 + \xi(r)) dV_{1} dV_{2} \, .
    \label{methodology:Eq:TPCFdP}
\end{equation}

\noindent In practice, for a catalogue of $ N $ particles and volume $ V $ with periodic boundary conditions, the 2PCF can be estimated by counting the number of pairs of objects $N_{\mathrm{s}}(r)$, in a shell of volume $V_{\mathrm{s}}(r)$, at distance $r$ from each other using

\begin{equation}
    \xi(r) = \frac{1}{n^{2}V} \frac{N_{\mathrm{s}}(r)}{V_{\mathrm{s}}(r)} - 1.
    \label{methodology:Eq:TPCFxi}
\end{equation}

\noindent However, if the boundaries are not periodic, this equation cannot be applied. This is the case  for a real survey, or when we restrict the analysis to a small region within a large periodic simulation. In these cases, the common method consists in comparing the observed data (D) with catalogues composed of random points (R) that reproduce the same geometry and artefacts of the original catalogue. In general, random catalogues contain at least 10 times more objects than the data catalogue, in order to reduce the noise level. In such cases, the 2PCF can be computed using the Landy-Szalay estimator \citep{Landy_Szalay_Estimator_1993}

\begin{equation}
    \xi(r) = \frac{DD(r) - 2 DR(r) + RR(r)}{RR(r)} \, .
    \label{methodology:Eq:LSEstimator}
\end{equation}

\noindent Here $ DD(r) $, $ DR(r) $ and $ RR(r) $ are the normalized data-data, data-random and random-random pairs, respectively. If $ N_{\mathrm{d}} $ and $ N_{\mathrm{r}} $ are the number of objects in D and R then

\begin{equation}
    DD(r) = \frac{dd(r)}{N_{\mathrm{d}} (N_\mathrm{d} - 1)/2} \, ,
    \label{methodology:Eq:DD}
\end{equation}
\begin{equation}
    RR(r) = \frac{rr(r)}{N_\mathrm{r} (N_\mathrm{r} - 1)/2} \, ,
    \label{methodology:Eq:RR}
\end{equation}
\begin{equation}
    DR(r) = \frac{dr(r)}{N_\mathrm{d} N_\mathrm{r}} \, ,
    \label{methodology:Eq:DR}
\end{equation}
\noindent where $ dd(r) $ is the number of objects pairs separated by a distance $ r $ in D, $ rr(r) $ is the number of pairs separated by a distance $ r $ in R and $DR(r)$ is the cross-correlation statistic, the number of pairs separated by a distance $r$ with one point taken from D and the other from R (for more details see e.g.~\citealt{Vargas_2013}). 
Throughout this work, we use the publicly available {\sc python} package {\sc corrfunc} \citep{Sinha_Corrfunc_2020} to compute correlation functions.

\begin{figure}
    \centering
    \includegraphics[width=0.9\linewidth]{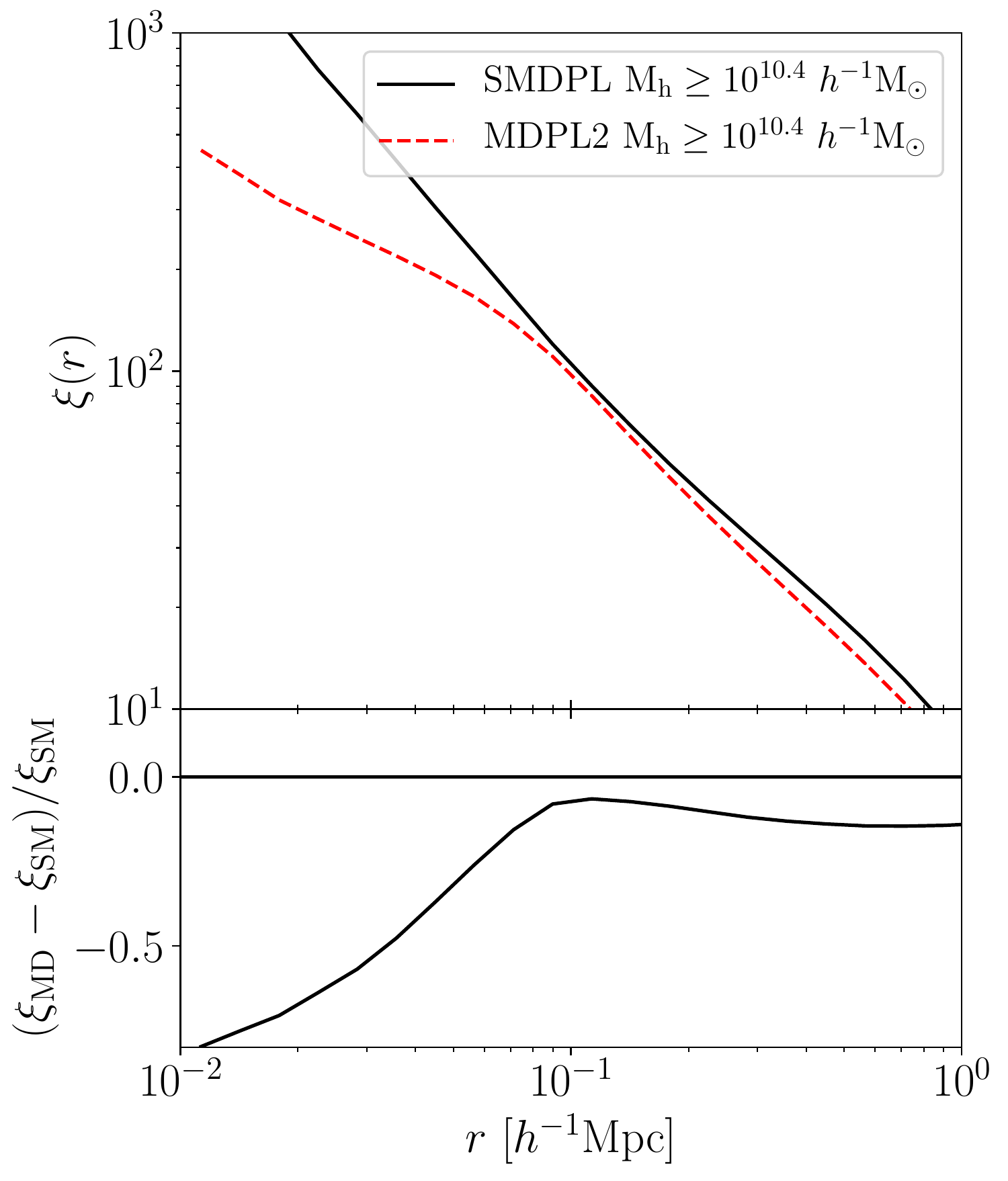}
    \caption{ \textit{Upper panel:} Two-point correlation function (2PCF) for the {\sc smdpl} (black solid line) and the {\sc mdpl2} (red dashed line) simulations at redshift $ z = 0 $. \textit{Lower panel:} The fractional difference between {\sc mdpl2} and {\sc smdpl} two-point correlation functions. For separations greater than $0.1 ~ h^{-1}\,\mathrm{Mpc}$, both simulations present a similar clustering. However, for smaller scales, the clustering functions differ considerably (fractional difference is greater than $ 0.5 $ for scales close to $0.01$). The lower subhalo number fraction in {\sc mdpl2}, compared with that in {\sc smdpl}, implies a lower number of pairs at low scales (where the 1-halo term dominates) and, therefore, a lower clustering at those scales.
    }
    \label{methodology:Fig:2PCF_MDPL2_SMDPL}
\end{figure}

Figure \ref{methodology:Fig:2PCF_MDPL2_SMDPL} shows the 2PCF for (sub)halo masses greater than $10^{10.4}~h^{-1}\,\mathrm{M_{\odot}}$ for the {\sc smdpl} (solid line) and {\sc mdpl2} (dashed line) simulations at redshift $ z = 0 $. From this figure, we see that the clustering of {\sc smdpl} is greater than that of {\sc mdpl2} for all scales; this effect is more significant at lower scales (between $0.01 - 0.1~h^{-1}\,\mathrm{Mpc}$). 
The suppression observed in the amplitude of the correlation function of {\sc mdpl2} could be due to the use of softened gravity or a finite mass resolution effect.
\cite{Jenkins_TPCF_softening_1998} shows that the softening length $ \varepsilon $ introduces considerable suppression in the clustering signal only for separations lower than $ 2 \varepsilon $. In our case, {\sc mdpl2} is characterised by $\varepsilon = 0.005 ~ h^{-1}\,\mathrm{Mpc}$, then for scales greater than $0.01~h^{-1}\,\mathrm{Mpc}$ the effect of the softening length should be very small. Therefore, the softening length cannot account for the large clustering suppression seen in the {\sc mdpl2} simulation for separations between $ 10^{-2} - 10^{-1}~h^{-1}\,\mathrm{Mpc}$.

According to the halo model, the 2PCF can be decomposed into two contributions: a 1-halo term and a 2-halo term. The 1-halo term involves correlations between haloes belonging to the same system, i.e. correlations between main host haloes and their corresponding 
subhaloes 
and correlations between all 
subhaloes 
that belong to a system. On the other hand, the 2-halo term involves correlations between haloes belonging to different systems \citep{Cooray_Sheth_HaloModels_2002}. When the contributions from these two terms are added together, the resulting correlation function should roughly follow a power law (see e.g. \citealt{Coil_LSS_Review_2013}). 
In general, the 2-halo term dominates at large scales (greater than $1~h^{-1}\,\mathrm{Mpc}$), while the 1-halo term dominates at scales lower than $1~h^{-1}\,\mathrm{Mpc}$. Note that the characteristic scale $\sim 1~h^{-1}\,\mathrm{Mpc}$, which is of the order of the size of the main systems, indicates the transition between the two regimes.
In \cite{vandenBosch_TPCF_LowScales_Satellites_2013}, it is shown that increasing the number of 
subhaloes 
increases mainly the 1-halo term, because subhaloes are located within host systems of typical sizes $\lesssim 1~h^{-1}\,\mathrm{Mpc}$. 
Therefore, the addition of USHs 
will enhance the clustering at small scales \citep{Kitzbichler_TPCFGalMergerRate_2008}.

For masses between $ 10^{10.4} - 10^{12}~h^{-1}\,\mathrm{M_{\odot}} $, the number fraction of 
subhaloes 
with respect to total 
(subhaloes 
plus main host haloes) is $ 0.150 $ for the {\sc smdpl} simulation, and $ 0.124$ for {\sc mdpl2}. This indicates that, for this mass range, we have roughly 20 per cent more 
subhaloes 
in {\sc smdpl} than in {\sc mdpl2}. Therefore, we conclude that the discrepancy observed between the 2PCFs of these simulations at small scales (see Figure \ref{methodology:Fig:2PCF_MDPL2_SMDPL}) is a result of the greater number fraction of 
subhaloes 
in {\sc smdpl} compared with {\sc mdpl2}.

To compensate for this lack of low-mass subhaloes that affects both the SHMF and 2PCF of the low-resolution {\sc mdpl2} simulation, we follow the dynamics of the USHs in {\sc mdpl2} in a semi-analytic way, by integrating their orbits until they merge with their host halo or are disrupted (see Section \ref{model}).
We assume that this procedure will provide better agreement with the SHMF and 2PCF provided by the high-resolution {\sc smdpl} simulation, which we consider adequate functions to constrain the free parameters of our orbital evolution model.

\begin{figure*}
    \centering
    \includegraphics[width=1.0\textwidth]{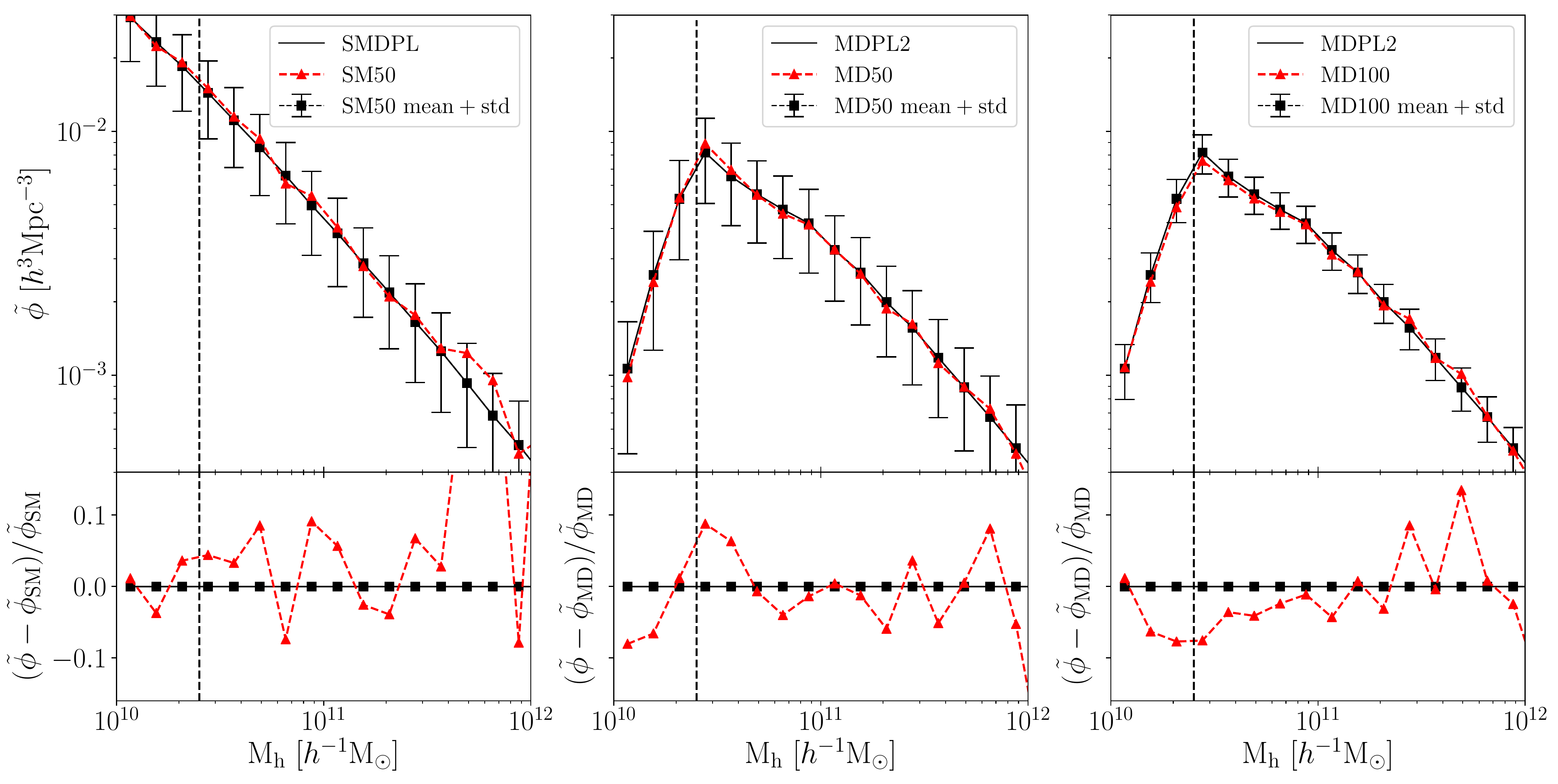}
    \caption{
    SHMF for the selected calibration and convergence-test boxes compared with those of the full simulations and the mean values obtained from all the 
    subvolumes. 
    Results are shown for the {\sc smdpl}  $50~h^{-1}\,\mathrm{Mpc}$ (SM50; left), {\sc mdpl2} $50~h^{-1}\,\mathrm{Mpc}$ (MD50; centre) and {\sc mdpl2} $100~h^{-1}\,\mathrm{Mpc}$ (MD100; right).
    \textit{Upper panels:} The continuous black 
    line represents the SHMF of the full box of the simulations. The square-dashed black line shows the mean value of the SHMF computed from the small boxes. The error bars correspond to the standard deviation at each subhalo mass bin. Red triangle-dashed lines correspond to the best boxes found to carry out the calibration and convergence-test procedures. \textit{Lower panels:} Fractional differences in SHMF taking the full simulations as a reference. The vertical line denotes the mass cut at $10^{10.4} ~ h^{-1}\,\mathrm{M_{\odot}}$. 
    }
    \label{methodology:Fig:HMF_boxes}
\end{figure*}

\subsection{Calibration and convergence-test  volumes}\label{methodology:Calibration}

\begin{figure*}
    \centering
    \includegraphics[width=1.0\linewidth]{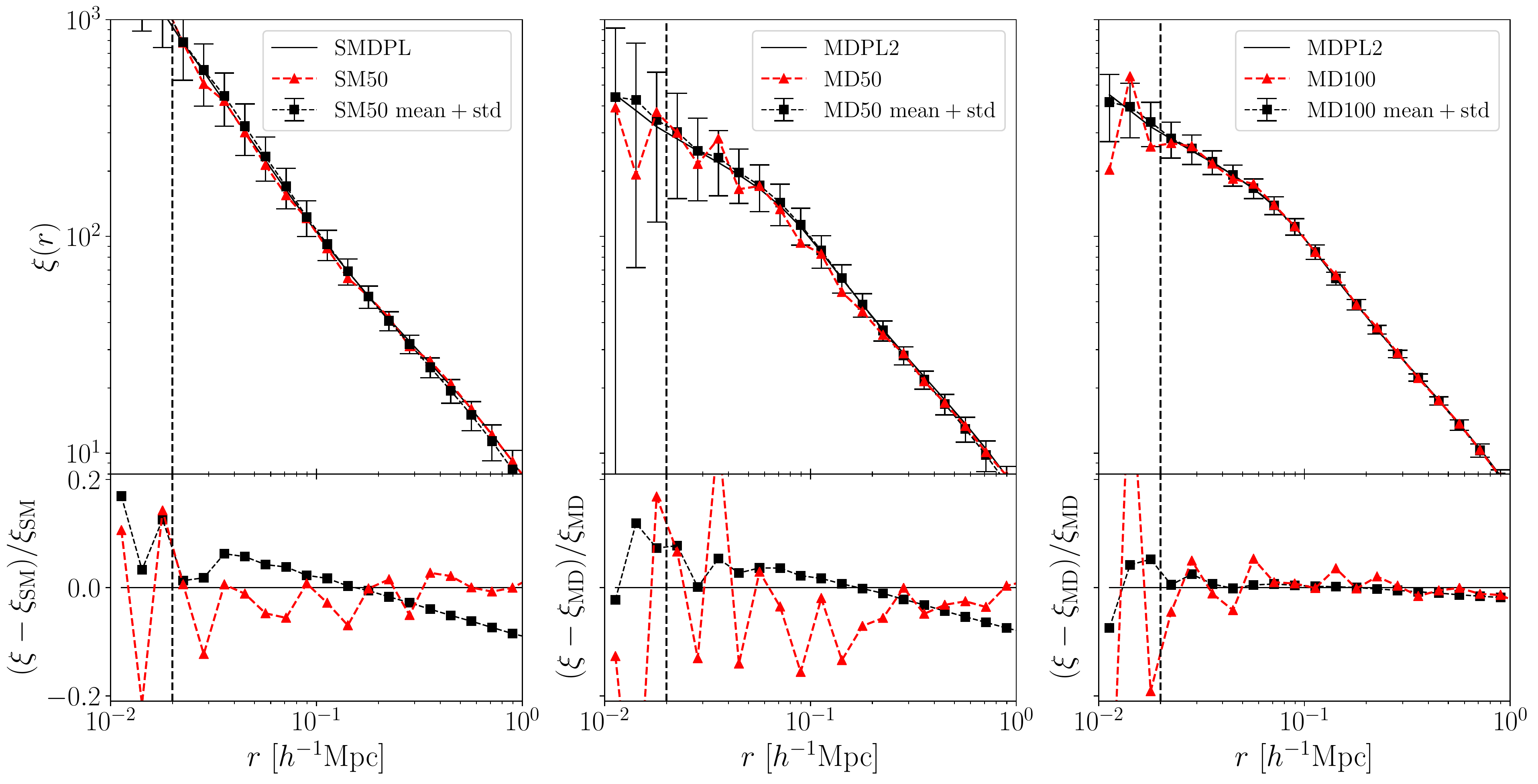}
    \caption{
    2PCF for the selected calibration and convergence-test boxes compared to the full simulations and the mean 
    values obtained from all the 
    subvolumes. 
    Results are shown for the {\sc smdpl}  $50~h^{-1}\,\mathrm{Mpc}$ (SM50; left), {\sc mdpl2} $50~h^{-1}\,\mathrm{Mpc}$ (MD50; centre) and {\sc mdpl2} $100~h^{-1}\,\mathrm{Mpc}$ (MD100; right). 
    \textit{Upper panels:} 
    The solid black line represents the full simulation. 
    The squared-dashed black line 
    shows the mean value for the 2PCF computed from the small boxes. The error bars correspond to the standard deviation. 
    The red triangle-dashed lines 
    correspond to the best boxes found to carry out the calibration and convergence-test procedures.
    \textit{Lower panels:} Fractional differences in 2PCF taking the full simulations as a reference. The vertical dashed line indicates the scale $0.2~h^{-1}\,\mathrm{Mpc}$. This limit corresponds to the smaller scale we use to make the comparison since, below this limit, boxes have very few pairs and the correlation function becomes too noisy. 
    }
    \label{methodology:Fig:2PCF_boxes}
\end{figure*}

The goal is to make a fast exploration of the parameters of the orbital evolution model, to find regions of the parameter space where there is convergence between the 
statistical properties obtained from simulations with different mass resolution (i.e., similar SHMF and 2PCF) after applying the orbital evolution model to the simulation with lower resolution.
This calibration procedure and the subsequent convergence test that we perform require running the model over both the low-resolution \textsc{mdpl2} and the high-resolution \textsc{smdpl} simulations. Since this task is computationally very expensive, we are interested in finding a set of boxes that are relatively small in volume but representative of the characteristics of the full simulations.
Then, we consider 
subvolumes 
of \textsc{mdpl2} with a box side length of  $50~h^{-1}\,\mathrm{Mpc}$ and $100~h^{-1}\,\mathrm{Mpc}$, and 
subvolumes 
of \textsc{smpdl} of $50~h^{-1}\,\mathrm{Mpc}$ on a side.
Thus, we have the subvolumes \textsc{mdpl2} $50~h^{-1}\,\mathrm{Mpc}$, \textsc{mdpl2} $100~h^{-1}\,\mathrm{Mpc}$ and  \textsc{smdpl} $50~h^{-1}\,\mathrm{Mpc}$, to which for simplicity we also refer to as MD50, MD100 and SM50, respectively. From the former set of \textsc{mdpl2} $50~h^{-1}\,\mathrm{Mpc}$ 
subvolumes 
we extract the best calibration box, while from the latter two sets we obtain the best convergence-test boxes. 
The choice of these best boxes requires the estimation of their ``goodness'' (capability of reproducing the characteristics of the full simulations). Here, we exemplify how to obtain the best calibration box, the other cases are analogous.

We partition the full box of {\sc mdpl2} into 8000 ($20^3$) disjoint subsamples with a box size of $50~h^{-1}\,\mathrm{Mpc}$. For each of these boxes, we calculate both $ \tilde{\phi} $ and $ \xi $. In the case of the 2PCF, $ \xi $, we consider only (sub)haloes with masses greater than $10^{10.4}~h^{-1}\,\mathrm{M_{\odot}}$, i.e. where the {\sc mdpl2} simulation is complete. Using these results, we compute the mean values of the SHMF and the 2PCF, $ \langle \tilde{\phi} \rangle $ and $ \langle \xi \rangle $, respectively. Analogously, we obtain the mean values $ \langle \tilde{\phi} \rangle $ and $ \langle \xi \rangle $ for both {\sc mdpl2} $100~h^{-1}\,\mathrm{Mpc}$ (with $10^3$ subvolumes) and {\sc smdpl} $50~h^{-1}\,\mathrm{Mpc}$ (with $8^3$ subvolumes).

Figures \ref{methodology:Fig:HMF_boxes} and \ref{methodology:Fig:2PCF_boxes} show (in square-dashed line) the mean values for the SHMF and 2PCF, respectively. We also plot (in continuous line) the SHMF and 2PCF corresponding to the full simulations. For both figures, we have the following cases: 
{\sc smdpl} $50~h^{-1}\,\mathrm{Mpc}$ (SM50 mean; left), {\sc mdpl2} $50~h^{-1}\,\mathrm{Mpc}$ (MD50 mean; centre) and {\sc mdpl2} $100~h^{-1}\,\mathrm{Mpc}$ (MD100 mean; right).  Error bars indicate the standard deviation of the 
subvolume 
sets. In Figure \ref{methodology:Fig:HMF_boxes}, we note that, as it happens in the {\sc mdpl2} full simulation, the boxes also present a lack of low mass haloes (centre and right). The vertical dashed line indicates the minimum mass we consider in our study ($10^{10.4}$~$h^{-1}$~$\mathrm{M_{\odot}}$). Figure \ref{methodology:Fig:2PCF_boxes} shows the clustering signal. 
We note that for very low separations ($ \lesssim 0.02 ~ h^{-1}\,\mathrm{Mpc}$), the scatter in $ \xi $ is high, which is reflected in the magnitude of the error bars. This effect is mainly due to the fact that, at these scales, the number of pairs is scarce and therefore Poissonian error dominates.

In order to find the best {\sc mdpl2} $50~h^{-1}\,\mathrm{Mpc}$ box (hereafter MD50), we need to characterise how representative each of our subsamples is. To estimate how much each box deviates from the full {\sc mdpl2} simulation (regarding to the SHMF and 2PCF), and pre-select some boxes that can be good candidates, we compute MAPE (mean absolute percentage error) estimates, i.e., for a given 
subvolume 
$j$, we compute the sum over all bins

\begin{equation}
    \mathrm{MAPE} \left[h^{(j)} \right] = \frac{1}{n} \sum^{n}_{i=1} \frac{ |h^{(j)}_{i} - h_{i}| }{ |{h}_{i}| }, 
    \label{methodology:Eq:MAPE_v2}
\end{equation}

\noindent where $h$ represents any of the relevant constraining functions (i.e., the $ \tilde{\phi} $ 
or $\xi$), $j$ indicates the given subvolume, 
and $ n $ is the number of bins  (mass-bins for $\tilde{\phi}$, separation-bins for $\xi$) where the function is computed. In this expression, $ {h}_i$ indicates the value of function $h$ in a given bin $i$ corresponding to the full simulation.

As mentioned above, the full {\sc mdpl2} simulation is not complete for low masses; furthermore, the correlation function computed in 
subvolumes 
becomes too noisy for scales below $0.02~h^{-1}\,\mathrm{Mpc}$. Taking this into account, for each 
subvolume 
we only consider $\mathrm{MAPE}\left[\tilde{\phi}\right] $ errors for masses higher than $10^{10.4}~h^{-1}\,M_{\odot}$ and $ \mathrm{MAPE}\left[\xi\right] $ errors for separations in the range $0.02 - 1 ~ h^{-1}\,\mathrm{Mpc}$. We choose candidates for the best MD50 box as the ones that simultaneously minimize both 
$ \mathrm{MAPE}\left[\tilde{\phi}\right] $ and $\mathrm{MAPE}\left[\xi\right] $. From these candidates, we choose our final best calibration box by visual inspection.

The method applied to find the best convergence-test 
subvolumes 
{\sc smdpl} $50~h^{-1}\,\mathrm{Mpc}$ (hereafter SM50) and {\sc mdpl2} $100~h^{-1}\,\mathrm{Mpc}$ (hereafter MD100) is analogous to what we have done for the calibration box MD50. 
Figure~\ref{methodology:Fig:HMF_boxes} shows in triangle-dashed red line the SHMF of the best 
subvolumes 
SM50 (left), MD50 (center) and MD100 (right); the SHMF of SM50 is slightly higher than the mean for masses close to $10^{12} ~ h^{-1}\,\mathrm{M_{\odot}}$.
Likewise, Figure~\ref{methodology:Fig:2PCF_boxes} shows in triangle-dashed red lines the 2PCF of the best 
subvolumes 
for the same sets. 
The vertical dashed lines in Figures~\ref{methodology:Fig:HMF_boxes} and~\ref{methodology:Fig:2PCF_boxes} indicate, respectively, the mass cut (for the SHMF; $10^{10.4} ~ h^{-1}\,\mathrm{M_{\odot}}$) and the smallest scale (for the 2PCF; $0.2~h^{-1}\,\mathrm{Mpc}$) we use to make the comparison with the corresponding mean values. In the case of the 2PCF, we can see that, below this limit, boxes have very few pairs and the correlation function becomes too noisy.
These results show that we are able to find smaller boxes, representative of the complete simulation boxes, to carry out the parameter exploration and the subsequent convergence test.

\section{Results}\label{results}

In this section, we make an exploration of the parameter space of the orbital evolution model of USHs applied to the low-resolution {\sc mdpl2} simulation by evaluating the changes produced on the SHMF and 2PCF when the free parameters of the model are varied.
We recall the meaning of the parameters under consideration: $b$ characterises the dynamical friction through the Coulomb logarithm (equation~\ref{model:Eq:LnCHashimoto2003}), $\alpha$ controls the TS (equation~\ref{model:Eq:TSrate}) and $f$ is related to the merger criterion (equation~\ref{model:Eq:f}).

\begin{figure*}
    \centering
    \includegraphics[width=0.9\linewidth]{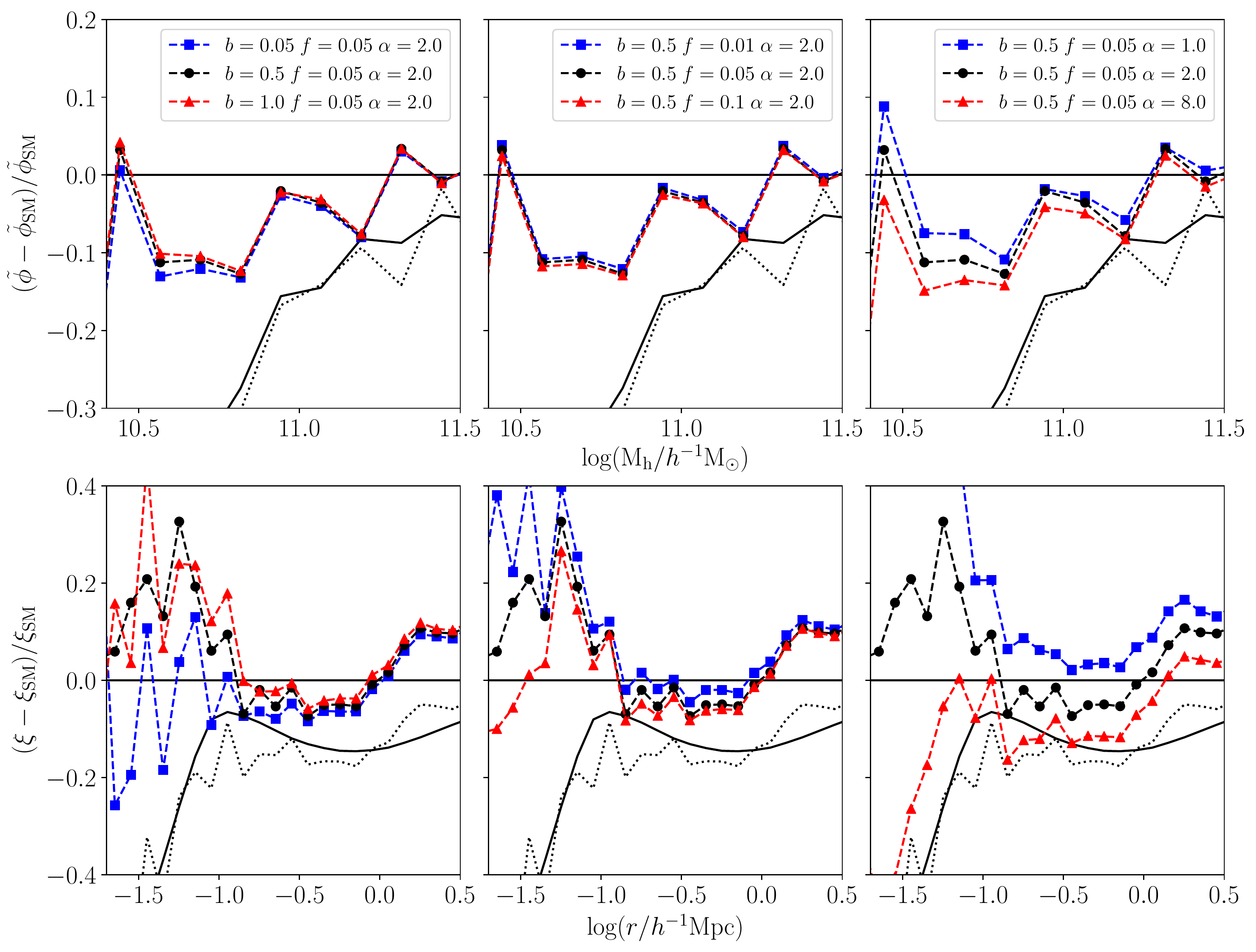}
    \caption{ 
    Fractional differences in SHMF (top panels) and in 2PCF (bottom panels) with respect to the {\sc smdpl} full simulation for a sample obtained from applying the orbital evolution model to USHs of the MD50 calibration box considering different combinations of values of the model parameters (dashed lined with symbols). The dashed-circle black line corresponds to the same fiducial parameters $(b = 0.50, f = 0.05, \alpha = 2.0)$ in all panels.
    Solid and dotted lines correspond to the fractional difference with respect to {\sc smdpl} for the {\sc mdpl2} simulation and the MD50 calibration box, respectively. 
    \textit{Left panels:} Effect of varying the parameter $ b $ leaving $ f $ and $ \alpha $ fixed. Parameter $ b $ enters the dynamical friction force term through the Coulomb logarithm, $ \ln({\Lambda}) $. Decreasing this parameter increases $ \ln({\Lambda}) $ and the DF effect, thus more USHs merge with their hosts and $ \tilde{\phi} $ decreases. Since we have a smaller number fraction of USHs, the clustering signal $ \xi $ also decreases. 
    \textit{Central panels:} Different values of the parameter $ f $ leaving $ b $ and $ \alpha $ fixed. The value of $ f $ determines the minimum distance an USH can be from the centre of its host before it is considered merged with it. Increasing this parameter implies a higher number of mergers and a lower number of USHs. This also impacts on the clustering signal $ \xi $, which decreases. 
    \textit{Right panels:} Effect of varying $ \alpha $ leaving  $ b $ and $ f $ fixed. Increasing the value of $ \alpha $ implies greater efficiency of the TS process and, consequently, greater mass loss which leads to a lower number of USHs with mass above the threshold $10^{10.4}~h^{-1}\,\mathrm{M_{\odot}}$ considered to estimate the SHMF and the 2PCF. As we have fewer USHs, then both $ \tilde{\phi} $ and $ \xi $ decrease.
    } 
    \label{results:Fig:parameter_variation}
\end{figure*}

Figure \ref{results:Fig:parameter_variation} shows the results obtained after applying the orbital evolution model to the calibration MD50 
subvolume 
introduced in Section \ref{methodology}. Here, we plot fractional differences taking the full {\sc smdpl} simulation as a reference, and selecting all (sub)haloes with masses greater than $10^{10.4}~h^{-1}\,\mathrm{M_{\odot}}$ at redshift $ z = 0 $. The top panels show fractional differences in the SHMF while the bottom panels show fractional differences in the 2PCF. The solid black line indicates the fractional difference between the {\sc mdpl2} and {\sc smdpl} full simulations, while the dotted black line indicates the fractional difference between the MD50 calibration box and the {\sc smdpl} full simulation. The dashed lines with symbols correspond to fractional differences obtained after applying the orbital evolution model to the MD50 calibration box, considering different parameter combinations of the model. In all panels, the dashed-circle black line corresponds to the same fiducial parameters $( b=0.50, f=0.05, \alpha=2.0)$.
From left to right, we vary $ b $ (left panels), $ f $ (centre panels) and $\alpha$ (right panels), leaving the remaining parameters fixed to their fiducial values.
In any case, when applying the orbital evolution model to the USHs, both the SHMF and 2PCF increase with respect to the results obtained from the {\sc mdpl2} simulation, as expected.
In the following subsections, we analyse the impact on the SHMF and the 2PCF of varying the different free parameters of our model.

\subsection{Variation of parameter $b$}
\label{results:Variation_b}

The parameter $ b $ enters the DF force term through the Coulomb logarithm (see equations \ref{model:Eq:Chandrasekhar1943} and \ref{model:Eq:LnCHashimoto2003}). Decreasing $ b $ is equivalent to increasing the value of the Coulomb logarithm $ \ln (\Lambda) $ and this leads to a greater deceleration of USHs due to DF drag and, consequently, a greater number of USHs that reach the central part of their host haloes. 
At first order, DF effect would only introduce changes in the spatial distribution of USHs, but mergers and TS may work in favour of modifying the number of USHs of a given mass. 
The impact of the changes in the parameter $b$ on the SHMF, $ \tilde{\phi} $, is shown in the upper-left  panel of Figure~\ref{results:Fig:parameter_variation} (dashed lines with symbols). Increasing (decreasing) $b$ results in a higher (lower) number of total USHs, and $ \tilde{\phi} $ increases (decreases). These changes are more noticeable for low-mass USHs. 
Although DF works more efficiently for more massive USHs, driving them to the inner parts of the host halo, the number of USHs that merge with their hosts is not high enough to produce a significant decrease in the SHMF. 
Indeed, for $b=0.05$, the lowest value of the parameter $b$ considered in this analysis, the number fractions of merged USHs\footnote{ This fraction is defined as the ratio between the number of merged USHs and the total number of unresolved systems at a given redshift; the latter includes the USHs that still survive according to the model, as well as  merged and disrupted USHs.
} (regardless of their mass) are 
$\sim 0.05$, $0.04$ and $0.01$ at $z = 2, 1, 0$, respectively. 
On the other hand, those USHs that do not merge are subjected to TS. This process becomes more efficient as the USHs move closer to the host centre (effect more pronounced for lower values of $b$). Therefore, when the mass of an USH is reduced by TS, it may be either disrupted or may fall below the mass cut considered in our analysis ($10^{10.4}~h^{-1}\,\mathrm{M_{\odot}}$), leading to a change in the SHMF at low masses (see further discussion in Section~\ref{results:Variation_alpha}).

The lower-left panel of Figure~\ref{results:Fig:parameter_variation} shows the impact of the changes in the parameter $b$ on the 2PCF, $ \xi $ (dashed lines with symbols).
The clustering signal increases (decreases) with the increase (decrease) of $b$ as a result of the increase (decrease) of the number fraction of USHs. The changes in $\xi$ are larger at small scales.
Since decreasing $ b $ implies that USHs decay faster towards the central part of their host halo, one might expect there should be an increase in clustering at lower scales. However, this does not occur because many of these USHs end up either merging with their main host halo or disrupting, and therefore do not contribute to the 2PCF.

\subsection{Variation of parameter $f$}\label{results:Variation_f}

The parameter $ f $ is related to the merger criterion. Increasing (decreasing) the value of $ f $ increases (decreases) the distance range within which USHs are considered to merge with their hosts.
Therefore, a higher value of $ f $ implies shorter merging time-scales, a greater number of mergers and a lower number of total USHs.
These trends affect both the SHMF and the 2PCF, as can be seen, respectively, in the upper-central and lower-central panels of Figure \ref{results:Fig:parameter_variation}, which show the results of the orbital evolution model for different values of the parameter $ f $. Thus, for a value of $f$ higher than the fiducial value, both the SHMF and the 2PCF decrease because of the lower number of USHs, while for a lower value of $ f $ both the SHMF and the 2PCF increase. However, the impact on the SHMF is minimal; the number of mergers is not significantly affected by the changes in $f$, and USHs of any mass are involved in the merging process.
For any value of $ f$, the value of the correlation function $\xi$ is larger for separations below $ \log (r/h^{-1} \mathrm{Mpc}) \leq -1 $ than at intermediate separations, i.e. $ \log (r/h^{-1} \mathrm{Mpc}) \geq -1 $.
Note that the increase in clustering signal obtained at low scales is more pronounced for a lower value of $f$ because the 1-halo term of $ \xi $ depends strongly on the number fraction of 
subhaloes, 
which is higher in this case.

\subsection{Variation of parameter $\alpha$} \label{results:Variation_alpha}

The parameter $ \alpha $ controls the rate at which the TS mechanism removes mass from an USH. The top-right and bottom-right panels of Figure \ref{results:Fig:parameter_variation} show, respectively (in dashed lines with symbols), the SHMF and the 2PCF for different values of $\alpha$. As the value of $\alpha$ increases, TS becomes more efficient and more material is removed from an USH via this process, 
leading to a reduction in the values achieved by the SHMF and the 2PCF. 
These trends are a result of the competition of two aspects. On one hand, the change of mass of the USHs as a result of TS have direct impact on their merging time-scales, i.e. a lower mass implies a lower dynamical friction force (equation \ref{model:Eq:Chandrasekhar1943}), leading to a larger number of USHs of lower mass that have not merged. 
On the other hand, some of the USHs subjected to mass removal might become disrupted leading to a decrease in the number fraction of USHs. Moreover, the mass of those USHs that have been neither disrupted nor merged may fall below the mass cut considered to estimate the SHMF and the 2PCF producing, respectively, a general decrease at all masses and scales. 
Indeed, after applying the orbital evolution model with the best-fitting parameters (see the following section), $\sim 90$ per cent of the USHs have masses below $10^{10.4}~h^{-1}\,\mathrm{M_{\odot}}$ at $z=0$.

\subsection{Best-fitting parameters}\label{results:BestFitting}

To find the best-fitting value of each free parameter, we perform an exploration of the parameter space running the orbital evolution model over the calibration box MD50 and varying the values of the parameters $ b $, $ f $ and $ \alpha $. For parameter $ b $, we consider values in the range $ 0.05 - 1.0 $. We allow $ f $ to vary between  $ 0.01 - 0.10 $. The parameter $ \alpha $ takes values between $ 1.0 - 8.0 $. 
For each run output, we compute the corresponding SHMF (for subhaloes with masses greater than $10^{10.4}~h^{-1}\,\mathrm{M_{\odot}}$) and the 2PCF (for haloes and subhaloes of masses greater than $10^{10.4}~h^{-1}\,\mathrm{M_{\odot}}$) at redshift $ z = 0 $. We then compare the values of these functions with those of {\sc smdpl} using a cost function  given by

\begin{equation} 
\begin{aligned}
    C(b, f, \alpha) = 
    & \frac{1}{N_{1}} \sum_{j=1}^{N_{1}} \ln^{2} \left( \frac{
    \tilde{\phi}^{j}_{\mathrm{model}}(b,f,\alpha)}{\tilde{\phi}^{j}_{\mathrm{\sc SM}}} \right) + \\
    & \frac{1}{N_{2}} \sum_{j=1}^{N_{2}} \ln^{2} \left( \frac{{\xi}^{j}_{\mathrm{model}}(b,f,\alpha)}{{\xi}^{j}_{\mathrm{\sc SM}}} \right),
    \label{results:Eq:cost_function} 
\end{aligned}
\end{equation}

\noindent where $ \tilde{\phi}^{j}_{\mathrm{\sc SM}} $ and ${\xi}^{j}_{\mathrm{\sc SM}} $ correspond to the SHMF and 2PCF of the {\sc smdpl} simulation, $\tilde{\phi}^{j}_{\mathrm{model}} $ and ${\xi}^{j}_{\mathrm{model}}$ are the corresponding SHMF and 2PCF that result from applying the orbital evolution model to the MD50 calibration box for a particular combination of the parameters $ b, f$ and  $ \alpha $, and $N_{1}$ and $N_{2}$ are the number of mass-bins and separation-bins used to estimate the calibration functions $ \tilde{\phi}^{j}_{\mathrm{\sc SM}} $ and ${\xi}^{j}_{\mathrm{\sc SM}} $, respectively.
Then, we select the best-fiting values of these parameters as those that minimize the cost function (equation  \ref{results:Eq:cost_function}). 
The most suitable parameters found with this method are $ b = 0.3$, $f = 0.036$, and $\alpha = 3.0 $.

\subsection{Convergence test}\label{results:ConvergenceTest}

\begin{figure}
    \centering
    \includegraphics[width=1.025\linewidth]{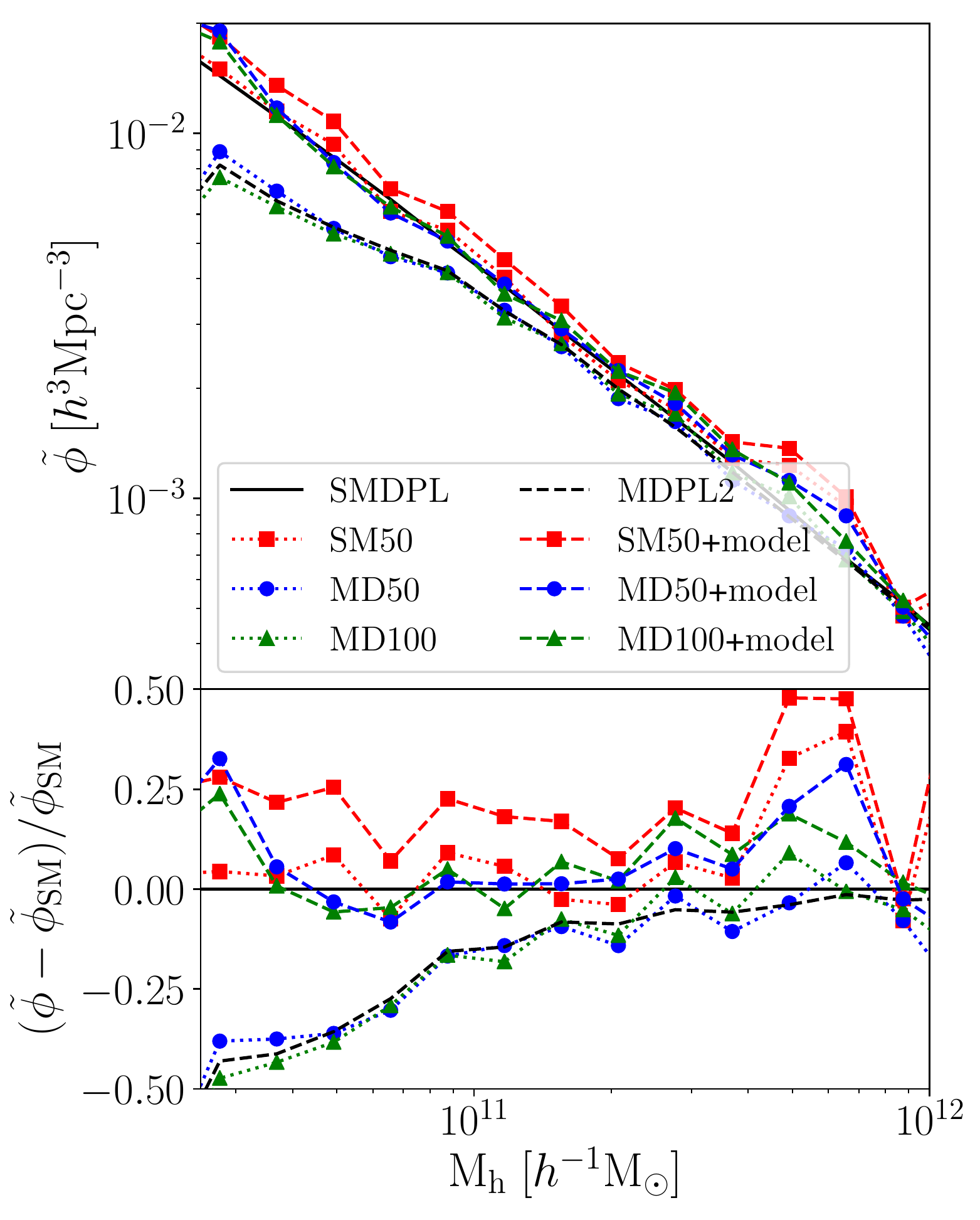}
    \caption{ 
    \textit{Upper panel:} Subhalo mass functions. Dotted lines with symbols correspond to calibration and convergence-test boxes (see Section \ref{methodology:Calibration}) without the inclusion of USHs. Dashed lines with symbols correspond to the those boxes after applying the orbital evolution model using the best-fitting parameters (see Section \ref{results:BestFitting}). We consider the \textsc{mdpl2} calibration box MD50, and the convergence-test boxes MD100 and SM50. The simple dashed line denotes the {\sc mdpl2} full simulation and the continuous line corresponds to the {\sc smdpl} full simulation. 
    \textit{Lower panel:} Fractional differences taking the {\sc smdpl} full simulation as a reference.  We see that after applying the model to track the orbital evolution of USHs in SM50 and MD50, the difference is 
    $\sim 10$ 
    per cent for the entire mass range. 
    }
    \label{results:Fig:BestFit_HMF}
\end{figure}

\begin{figure}
    \centering
    \includegraphics[width=1.0\linewidth]{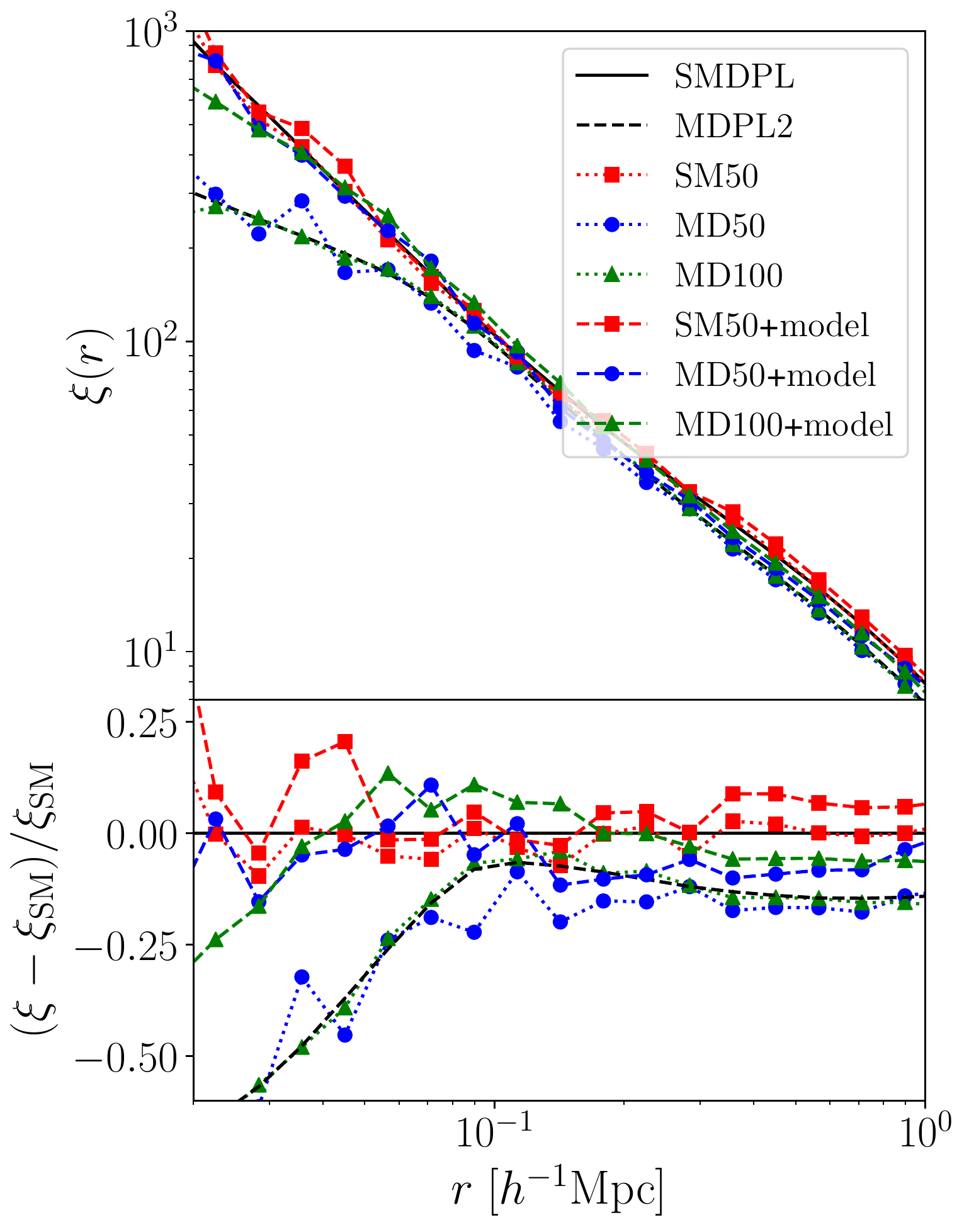}
    \caption{ 
    Two-point correlation functions. Dotted lines with symbols correspond to the calibration and convergence-test boxes without the inclusion of USHs. Dashed lines with symbols correspond to the samples obtained for these boxes after applying the orbital evolution model using the best-fitting parameters. We consider the calibration box MD050, and the convergence-test boxes MD100 and SM050. The simple dashed line denotes the full {\sc mdpl2} simulation and the continuous line corresponds to the full {\sc smdpl} simulation. 
    \textit{Lower panel:} Fractional differences in the 2PCFs taking the full {\sc smdpl} simulation as a reference. After applying the model, we obtain an enhancement of the clustering for both MD050 and MD100. In general, the agreement between the different boxes is within 10 per cent except for scales below $0.03~h^{-1}\,{\rm Mpc}$, where the agreement is within 20 per cent.
    }
    \label{results:Fig:BestFit_2PCF}
\end{figure}

We conduct a convergence test by comparing results obtained for the calibration box MD50 with those for the convergence-test boxes, MD100 and SM50, introduced in Section \ref{methodology:Calibration}. Such comparison involves results obtained by considering only resolved subhaloes, on one hand, and by adding the USHs, on the other. Comparisons between MD50 and MD100 with and without USHs are done to evaluate the impact of the box size on the results obtained. Comparisons of results for MD50 with and without USHs with those obtained for the full \textsc{smdpl} simulation and the SM50 
subvolume 
have the purpose of validating the orbital evolution model and its best-fitting parameters.
Thus, we run the orbital evolution model using the best-fitting parameters over the boxes MD50, MD100 and SM50.
The results of this exercise are shown (in dashed lines with symbols) in Figure \ref{results:Fig:BestFit_HMF} and Figure \ref{results:Fig:BestFit_2PCF} for the SHMF and 2PCF, respectively.
The results for the  boxes without the inclusion of USHs are depicted as dotted lines with symbols.
As a reference, we plot the \textsc{mdpl2} full simulation (in dashed line) and the \textsc{smdpl} full simulation (in continuous line). In the lower panel of these figures, we show fractional differences taking the {\sc smdpl} full simulation as a reference.

From Figure \ref{results:Fig:BestFit_HMF} we can see that, after applying the orbital evolution model to follow USHs, a good agreement with the {\sc smdpl} full simulation is achieved for the different boxes (dashed lines with symbols), compared with the behaviour obtained considering only the detected subhalos in those boxes (dotted lines with symbols). 
We also notice that both MD50 and SM50 present some ``spikes'' at masses of the order of $10^{12}~h^{-1}\,\mathrm{M_{\odot}}$; this is a particular characteristic of the selected boxes. The inclusion of USHs in SM50 (dashed line with squares) has little impact on the SHMF.
This is quantified by estimating the number density of subhaloes for the mass range above the mass cut considered ($10^{10.4} ~ h^{-1}\,\mathrm{M_{\odot}}$): $\sim 0.0172 ~ h^{3}\,\mathrm{Mpc}^{-3}$ for SM50, and 
$\sim 0.0201 ~ h^{3}\,\mathrm{Mpc}^{-3}$ 
for SM50 with USHs (SM50$+$model).
For MD50, we note that the addition of USHs (dashed line with circles) helps to enhance the SHMF at low masses; the number density of all resolved subhaloes in MD50 is $\sim 0.0121 ~ h^{3}\,\mathrm{Mpc}^{-3}$, and the number density obtained after adding USHs increases to 
$\sim 0.0177 ~ h^{3}\,\mathrm{Mpc}^{-3}$, 
achieving a very good agreement with the value obtained for SM50. Note that these number densities depend on the chosen
subvolumes. 
Differences between MD50 and SM50 also arise because {\sc mdpl2} not only presents a deficiency in subhaloes but also of main host systems compared with {\sc smdpl}.
The SHMF obtained for MD50 after adding UHSs (MD50$+$model) converges to the SHMF of the full \textsc{smdpl}; the mean fractional difference over the entire mass range is 
$\sim 10$ 
per cent. Finally, the results obtained for both MD100 (dotted line with triangles) and MD100$+$model (dashed line with triangles) are very similar to those obtained for MD50 and MD50$+$model, respectively, indicating that the size chosen for the calibration box is adequate to carry out the calibration procedure.

Figure \ref{results:Fig:BestFit_2PCF} shows the effect on the 2PCF of running the orbital evolution model over the calibration and convergence-test boxes using the best-fitting parameters.
In general, tracking the orbital evolution of USHs in MD50 and MD100 (MD50$+$model and MD100$+$model, respectively) considerably improves the 2PCF, increasing the signal over the entire range of scales (dashed lines with symbols) compared to the results obtained considering only the detected substructures in those boxes (dotted lines with symbols). 
Note that for both {\sc mdpl2} subvolumes in the latter case, the clustering is strongly suppressed with respect to the one obtained for the full {\sc smdpl} simulation with a discrepancy greater than 50 per cent for scales close to $0.02~h^{-1}\mathrm{Mpc}$.
The enhancement obtained when including USHs is more pronounced at small scales ($0.02 - 0.10 ~ h^{-1}\,\mathrm{Mpc}$) where the 1-halo term dominates, which mainly depends on the number fraction of satellite haloes. 
In general, mean fractional errors for the 2PCF over the whole range of scales considered are within 10 per cent for all calibration and convergence-test boxes (i.e. MD50, MD100 and SM50) when adding USHs.
From these results, we can conclude that the inclusion of USHs improves the SHMF and 2PCF of the simulation with lower resolution ({\sc mdpl2}) leading to a better agreement with the results obtained for the higher-resolution one ({\sc smdpl}).

With this analysis, we have shown that, on the one hand, the {\sc smdpl} simulation has 
an almost 
complete population of subhaloes above $10^{10.4}~h^{-1}\,\mathrm{M_{\odot}}$.
Indeed, if we apply the orbital evolution model to a smaller fraction of {\sc smdpl}, namely SM50, we obtain almost the same SHMF and 2PCF as in the full simulation. 
The differences between {\sc smdpl} and its SM50 box after the application of the model are a result of the presence of unresolved systems in {\sc smdpl}, and are only noticeable at the smallest masses (in the SHMF) and smallest scales (in the 2PCF); these differences are small and we can safely neglect them.
This fact supports the use of the {\sc smdpl} simulation as a reference for the exploration of the parameters of the orbital evolution model. 
On the other hand, the results obtained for the MD50 and MD100 boxes after including USHs are consistent (the observed differences are a result of the selected box). Clearly, the size of the MD50 box is sufficiently good to tune the free parameters of the model, and it is not necessary a larger box for this purpose.


\section{Validation of the calibration procedure} \label{results:ValidationProcedure}

\noindent Recently, \cite{vdBosch_Ogiya_2018} investigated the circumstances in which subhaloes identified in cosmological simulations undergo disruption, arriving at the conclusion that most disruptions are numerical in origin. Using a large suite of idealized simulations in which subhaloes move in a fix, external potential, they analyse under what conditions inadequate force softening and discreteness noise have an impact on subhaloes.
In their work, the authors present two criteria (applied to the bound mass fraction $f_{\rm bound}$ of subhaloes) to assess whether individual subhaloes in cosmological simulations are reliable or not.
In our work, we estimate $f_{\rm bound}$ as the ratio between the mass of the subhalo at a given time and its mass at accretion.
They claim that the subhaloes that satisfy either of these two criteria should be discarded from further analysis. 
Specifically, they find that subhaloes  start to be significantly affected by discreteness noise when

\begin{equation}
    f_{\rm bound} < 0.32 \left( \frac{N_{\rm acc}}{1000}\right)^{-0.8},
\label{appendixB:Eq:fbound1} 
\end{equation}

\noindent where $N_{\rm acc}$ is the number of particles in the subhalo at accretion.
Furthermore, subhalos are systematically affected by inadequate force resolution when

\begin{equation}
    f_{\rm bound} < \frac{1.79}{f(c)} \left( \frac{\varepsilon}{r_{\mathrm{s},0}}\right) \left( \frac{ r_{\rm h}}{r_{\mathrm{s},0}} \right),
\label{appendixB:Eq:fbound2} 
\end{equation}

\noindent where $r_{\mathrm{s},0}$ and $c$ are the scale radius and concentration parameter of the (NFW) subhalo at accretion, respectively, $ f(x) = \ln{(1 + x)} - x/(1 + x) $, $\varepsilon$ is the characteristic softening length of the simulation and $r_{\mathrm{h}}$ is the instantaneous half-mass radius.
Both situations could lead to artificial disruptions and, therefore, they are considered as criteria to discard subhaloes.

The authors indicate that DM subhaloes are artificially disrupted even in state-of-the-art cosmological {\em N}-body simulations. Therefore, one could argue that the  {\sc smdpl} simulation, which is taken as reference in the present work, may be subjected to numerical artifacts that affect the tidal evolution of the DM subhaloes. 
If we apply those criteria to the cosmological simulations that we are using in our analysis, most of the subhaloes would not survive. In fact, this would be true for many of the available simulations in the literature, since those findings are applicable to all the subhaloes identified in any cosmological simulations.
However, the results of any study that analyse the convergence between simulations computed with similar techniques will still be valid and useful for modelling structure formation.

In our case, we are studying the convergence between two simulations ({\sc mdpl2} and {\sc smdpl}) with different mass and spatial resolutions that have been selected from the same family (MultiDark-Planck simulations), i.e., they  have been generated using the same methodology and cosmology, and have been analysed with the same halo finder ({\sc rockstar}). Therefore, they suffer similar numerical artifacts.
The method that we use to select the parameters that best fit our orbital evolution model, which is applied to the USHs of the lower-resolution simulation ({\sc mdpl2}), tries to find the convergence of some global statistical properties (i.e. SHMF and 2PCF) between both simulations.

To assess the robustness of the calibration procedure used to determine the best-fitting parameters of the orbital evolution model presented in this work, we perform a comparison between the orbits followed by a set of resolved subhaloes in the  SM50 box (\textsc{smdpl} $50~h^{-1}\,\mathrm{Mpc}$, see section \ref{methodology:Calibration}), that are reliable according to the criteria presented in \cite{vdBosch_Ogiya_2018} (i.e. they do not satisfy neither equation~(\ref{appendixB:Eq:fbound1}) nor equation~(\ref{appendixB:Eq:fbound2})), and the orbits of those same resolved subhaloes predicted by our orbital evolution model.
Such comparison is quantified by analysing the distribution of values adopted by the free parameters of the model. The aim is to evaluate if they are consistent with the values of the best-fitting parameters found using the SHMF and 2PCF as constraints (section~\ref{results:BestFitting}).

In this analysis, we consider the circularity of bound orbits (negative total energy, $ E < 0 $).
The circularity is defined as the ratio of the orbital angular momentum, $L$, to the angular momentum of a circular orbit with the same energy, $ L_{\mathrm{\mathrm{circ}}}(E) $. Hence, the circularity of an orbit is given by

\begin{equation} 
    \eta = \frac{L}{ L_{\mathrm{\mathrm{circ}}}(E) } = \frac{L}{ r_{\mathrm{circ}} \, \mu \, V_{\mathrm{circ}} },
    \label{appendixB:Eq:circularity} 
\end{equation}

\noindent where 
$ r_{\mathrm{circ}} = - G M_{\mathrm{host}} M_{\mathrm{sub}} / E $,
$ V_{\mathrm{circ}} = \sqrt{ -2 E / \mu}$, 
and 
$ \mu = M_{\mathrm{host}} M_{\mathrm{sub}} / (M_{\mathrm{host}} + M_{\mathrm{sub}}) $ 
is the reduced mass of the 
host-subhalo system 
\citep{Khochfar_Burkert_OrbitalParameters_2006}.

To carry out the comparison, we select reliable subhaloes (according to \cite{vdBosch_Ogiya_2018} criteria) that survive in the SM50 box for a minimum of $ 25 $ snapshots. For each subhalo, we compute the circularity along its orbit according to equation \ref{appendixB:Eq:circularity}. We further separate the subhaloes in bins of circularity, keeping only those with stable circularity (i.e. those for which its circularity at different snapshots does not deviate more than 0.15 from its mean circularity estimated during the period of time in which we compare the actual orbit with the one simulated by our model).
We run the orbital evolution model for these subhaloes and estimate separately the values of the parameter $b$, involved in the DF model  (Section~\ref{results:Variation_b}), and the parameter $\alpha$, associated to TS (section~\ref{results:Variation_alpha}). 
We do not estimate  the parameter involved in the proximity merger criterion, $f$ (section~\ref{results:Variation_f}), because several subhaloes in the sample selected for this analysis remain resolved up to $z=0$, that is, they do not merge with their host. We then proceed in the following way:

\begin{enumerate}
    \item Firstly, we estimate the value of the parameter $b$ 
    by running 
    the orbital evolution model 
    with different values of this parameter. 
    Although DF affects both the positions and velocities of the subhaloes, we focus on the former. 
    In order to avoid the impact of the TS modelling, the mass and radius of the subhalo 
    ($M_{\rm ush}$ and $R_{\rm ush}$
    in equations \ref{model:Eq:Chandrasekhar1943} and \ref{model:Eq:LnCHashimoto2003}, respectively) 
    are assigned according to the values that those quantities take along the orbit in the {\sc smdpl} cosmological simulation.
    Then, we compare the evolution of the positions of the subhaloes in our sample (taken from the SM50 box) with that predicted by the orbital evolution model (equations \ref{model:Eq:Chandrasekhar1943} and \ref{model:Eq:LnCHashimoto2003}). 
    
    \item Secondly, we estimate the value of the parameter $ \alpha $ involved in the TS model, which affects the mass of the subhaloes, by varying this parameter in the orbital evolution model. In this case, we do not consider the DF modelling, and the positions and velocities of the subhaloes are extracted from the {\sc smdpl} cosmological simulation. 
    Then, we compare the evolution of the mass of the subhaloes in our sample (taken from the SM50 box) with the evolution of the mass as predicted by our model (equations \ref{model:Eq:TidalRadiusKing} and \ref{model:Eq:TSrate}). 
    
\end{enumerate}

To quantify the similarity of the actual orbit of the subhaloes with the orbit simulated by our orbital evolution model, we construct a cost function that is the sum of the logarithmic squared differences between these two types of orbits along the time.
We use many orbits to perform the analysis, and average the cost functions of all of them as our final cost function $C(p)$. This cost function depends on the parameters $p$ of the orbital evolution model (either $b$ or $\alpha$), and its minimum indicates which is the model parameter for which the likeness is the best. Hence, the cost function is

\begin{equation} 
    C(p) =  \sum_{i=1}^{N_{\mathrm{sim}}} \frac{C_{i}(p)}{N_{\mathrm{sim}}},
    \label{appendixB:Eq:cost_function_av} 
\end{equation}

\noindent where $C_i(p)$ is the cost function for each orbit, 
which depends on the parameter $p$, and $ N_{\mathrm{sim}} $ is the number of orbits included in the analysis. In this analysis we consider $N_{\mathrm{sim}} = 25$.

\noindent For the parameter $b$, the cost function of a single orbit is

\begin{equation} 
    C_i(b)   =  \frac{1}{N^{i}_{\mathrm{evals}}} \sum_{j=1}^{N^{i}_{\mathrm{evals}}} \ln^{2} \left( \frac{r^{i}_{\mathrm{model}}(t_j , b)}{r^{i}_{\mathrm{\sc SM50}}(t_j)} \right).
    \label{appendixB:Eq:cost_function_b} 
\end{equation}

\noindent Here, 
$ r^{i}_{\mathrm{model}}(t_j, b) $ is the radial distance from the subhalo to the centre of the host 
obtained from the orbital evolution model, where $b$ is the varied parameter and $ t_j $
is the point of evaluation (in the orbit);
$ r^{i}_{\mathrm{\sc SM50}}(t_j) $ is the value of radial distance along the actual orbit extracted from the SM50 box. 
Since the selected subhaloes have different lifetimes,
the number of evaluations, $ N^{i}_{\mathrm{evals}} $, depends on the orbit considered.

\noindent For the parameter $\alpha$, the cost function for a single orbit is

\begin{equation} 
    C_i(\alpha) = \frac{1}{N^{i}_{\mathrm{evals}}} \sum_{j=1}^{N^{i}_{\mathrm{evals}}} \ln^{2} \left( \frac{m^{i}_{\mathrm{model}}(t_j , \alpha)}{m^{i}_{\mathrm{\sc SM50}}(t_j)} \right) .
    \label{appendixB:Eq:cost_function_alpha} 
\end{equation}

\noindent In this case, we compare the  mass of the subhalo 
provided by the {\sc smdpl} simulation (SM50 box), $ m^{i}_{\mathrm{\sc SM50}}(t_j) $, with that
computed by applying the orbital evolution model, $m^{i}_{\mathrm{model}}(t_j, \alpha) $, where $\alpha$ is the varied parameter and $ t_j $ is the point of evaluation (in the orbit).

\begin{figure}
    \centering
    \includegraphics[width=0.9\linewidth]{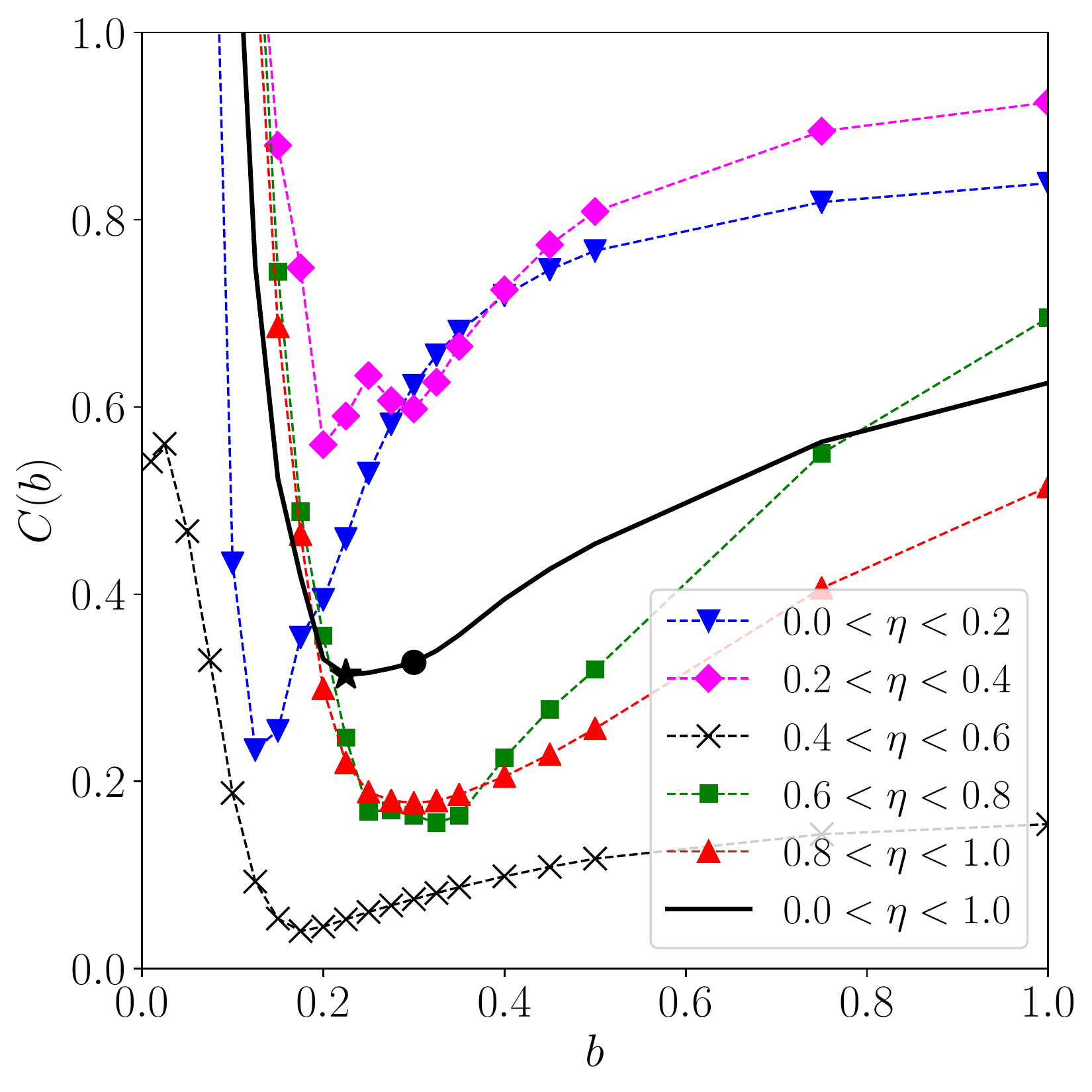}
    \caption{
    Cost function for the parameter $b$ for subhaloes selected from the SM50 box. Dashed lines with symbols correspond to subhaloes lying within different circularity bins, while the continuous line represents the full sample of  subhaloes. The value of $b$ that minimizes the cost function depends on the circularity of the orbit.
    The black-circle represents the cost function for the best-fitting parameter $ b = 0.3$. The black-star symbol, located at $ b = 0.225 $, indicates the minimum value of the cost function for the sample of subhaloes analysed here.
    }
    \label{appendixB:Fig:Cost_b}
\end{figure}

Figure \ref{appendixB:Fig:Cost_b} shows the cost function for different values of $ b $, both for subhaloes contained in different circularity bins (dashed lines with symbols) and for the full sample of selected haloes (continuous line).
We note that the value of $ b $ that minimize the cost function depends on the value of the circularity of the orbits.
For the complete sample considered for this analysis, we obtain that a value of $b= 0.225$ (black-star symbol) optimise the similarity between the orbits of the resolved subhaloes in the SM50 box and the corresponding orbits predicted by the orbital evolution model.

\begin{figure}
    \centering
    \includegraphics[width=0.9\linewidth]{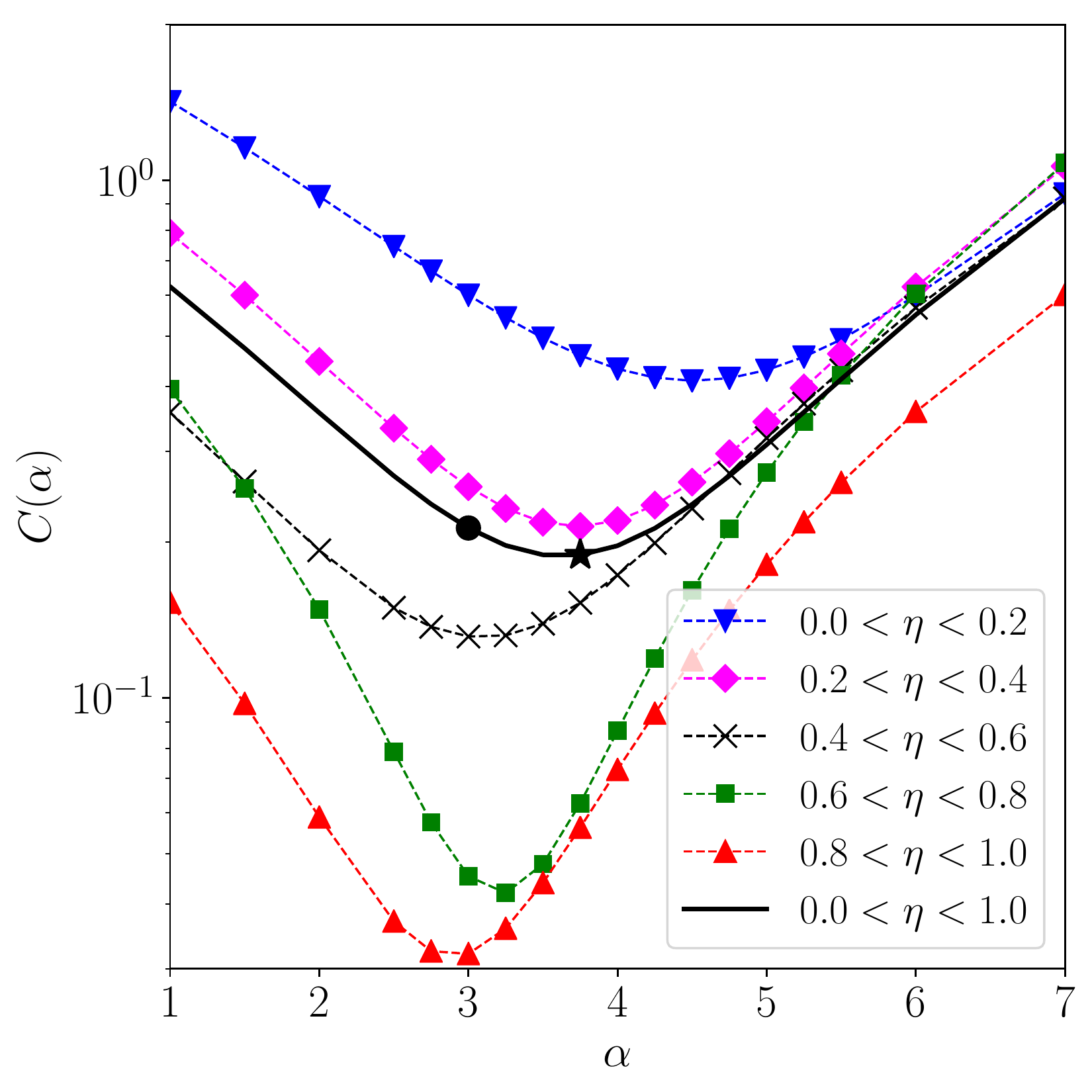}
    \caption{
    Cost function for the parameter $\alpha$ for subhaloes selected from the SM50 box. Dashed lines with symbols correspond to subhaloes lying within different circularity bins, while the continuous line represents the full sample of subhaloes. The value of $\alpha$ that minimizes the cost function depends on the circularity of the orbit. 
    The black-circle indicates the cost function for the best-fitting parameter $ \alpha = 3.0 $. The black-star symbol, located at $ \alpha = 3.75 $, corresponds to the minimum value of the cost function for the sample of subhaloes analysed here.
    }
    \label{appendixB:Fig:Cost_alpha}
\end{figure}

Figure \ref{appendixB:Fig:Cost_alpha} shows the cost function for different values of the parameter $\alpha $. In dashed lines with symbols, we show the cost function for subhaloes lying in different circularity bins, while the continuous line corresponds to the cost function of the full sample. The parameter $\alpha$ also depends on the value of the circularity of the orbit, showing a clear trend. This is reasonable, since for more circular orbits (large values of $\eta$), the subhalo will not reach the innermost part of the host gravitational potential and, hence, it will be less affected by TS  (lower values of $\alpha$). 
On the contrary, less circular orbits (more elongated ones) will reach the innermost parts of the host potential, being subjected to a larger tidal stripping effect, resulting in larger values of the parameter $\alpha$. 
When using the complete sample of selected subhaloes, the cost function is minimized for $\alpha = 3.75$ (black-star symbol).

The value of the cost functions for the best fitting parameters $b = 0.3$ and $\alpha = 3.0$, are represented by black-circles in Figures \ref{appendixB:Fig:Cost_b} and \ref{appendixB:Fig:Cost_alpha}, respectively. Note that these values are consistent with the values $b$ and $\alpha$ that we get when analysing individual orbits of reliable subhaloes, according to the criteria of \cite{vdBosch_Ogiya_2018}, 
that is, by using the SHMF and the 2PCF to constrain the free parameters of the orbital evolution model. Moreover, our methodology, which involves such global statistics, naturally takes into account, and average out, the different behaviours that we see for different circularities. 
Hence, the results of this analysis give strong support to the calibration procedure applied to our orbital evolution model.

\section{Discussion}\label{discussion}

We present a model to track the orbital evolution of USHs and propose a method to calibrate its free parameters.
From the analysis presented above we have that, in general, the inclusion of USHs in the statistical analysis of the properties of substructures (e.g. mass and phase-space coordinates) detected in a DM-only simulation is key to achieving a better agreement with the results obtained from a higher-resolution simulation in which the number of USHs is smaller. Here, we discuss in more detail some aspects of our results.

The inclusion of USHs helps to enhance the SHMF over the entire mass range.
However, as we can see from the upper-panels of Figure \ref{results:Fig:parameter_variation}, the SHMF depends more strongly on the parameter $ \alpha $ (related to TS, which changes the mass of the USH), than on the parameters $ b $ (involved in DF) and $ f $ (considered in the merger proximity criterion). Indeed, varying $b$ and $f$ has little impact on the overall shape and normalisation of the SHMF. Since $ b $ and $ f $ are related mainly to the evolution of the position of the USHs, the SHMF fails to put a tight constraint on those parameters.

The main contribution of the calibration procedure proposed in this work is the inclusion of the 2PCF as a constraining function, which is sensitive to variations of the three parameters involved in our orbital evolution model (see lower panels of Figure \ref{results:Fig:parameter_variation}), although the behaviour is different depending on the scales considered. For very small scales ($ \lesssim 0.3~h^{-1}\,\mathrm{Mpc} $), the correlation function is dominated by the 1-halo term (\citealt{vandenBosch_TPCF_LowScales_Satellites_2013}) which depends strongly on the number fraction of 
subhaloes. 
Then, this region is very sensitive to variations of the parameters involved in the dynamical friction model ($ b $ and $\alpha$ through the Coloumb Logarithm and the 
subhalo 
mass, respectively) or the merging criterion ($ f $). 
At greater separations ($ \gtrsim 0.3 ~ h^{-1}\,\mathrm{Mpc} $), the 2-halo term begins to compete against the 1-halo term, and the constraining power of the correlation function is reduced. It is worth noting that if the strength of TS is high ($\alpha \sim 8.0$), then the clustering is strongly suppressed at small scales. Hence, while the SHMF fails to constrain $ b $ and $ f $, the 2PCF has great constraining power on all free parameters of the model.
As a result of the exploration of the parameter space considering the SHMF and 2PCF as constraints, we find the following best-fitting parameters: $ b=0.3, f=0.036, \alpha=3.0$.

Previous works in the literature explore the possible values that the Coulomb logarithm can take. \cite{Pullen_SubhaloNonlinearEvolution_2014} assume a fixed value $ \ln (\Lambda) = 2 $. Other authors choose the value of the Coulomb logarithm in order to reproduce the results of numerical simulations. For example, \cite{Velazquez_White_DynFriction_1999}, by studying orbits in {\em N}-body simulations, find values of $\ln (\Lambda)$ within the range $1 - 2$.  Using a model similar to the one presented here, \cite{Yang_SubHaloMCMC_2020} find $ \ln (\Lambda) \sim 1.5 $ from the parameter space exploration. 
In our case, for the best-fitting parameter $b=0.3$, the range of values covered by the Coulomb logarithm implemented in the orbital evolution model (equation~\ref{model:Eq:LnCHashimoto2003}) is $0 - 7$.
The minimum value, $ \ln (\Lambda) = 0 $, indicates that the DF term can only produce deceleration, and is adopted when 
$ r_{\rm ush} \leq b R_{\rm ush} $. 
Clearly, the maximum value of the Coulomb logarithm we obtain is larger than those found in previous works, but it is consistent with a broad estimation based on the typical values of the virial radius of haloes (and subhaloes). In our simulations, these radii are in the range $10 - 1000~h^{-1}\,\mathrm{kpc}$. Then, assuming a maximum satellite-host centre distance 
$r_{\rm ush}$ 
of the order of the virial radius of main systems, we have that, at most, 
$ r_{\rm ush}/R_{\rm ush} \sim R_{\rm host}/R_{\rm ush} \sim 100 $. 
This gives an estimated maximum value of the Coulomb logarithm of $ \ln (\Lambda)_{\mathrm{max}} \sim  6.2$, consistent  with that obtained when applying the orbital evolution model. 
In the latter case, the mean value of the distribution of Coulomb logarithm at $z=0$ is $ \ln (\Lambda)\sim 3.6$.

The discrepancy between the mean value of the Coulomb logarithm we have derived and those reported by \cite{Velazquez_White_DynFriction_1999}, \cite{Pullen_SubhaloNonlinearEvolution_2014} and \cite{Yang_SubHaloMCMC_2020} could be attributed to several reasons.
On one hand, the results of these works were obtained by adopting a constant value for $ \ln (\Lambda) $, whereas in our orbital evolution model we assume a Coulomb logarithm that varies with the distance of the USH to the host centre.
On the other hand, apart from DF, the deepening of the gravitational potential of the host contributes to shrinking the orbits of subhaloes. 
The potential well of haloes increases as they grow by a combination of relatively smooth accretion and mergers with smaller structures.
\cite{Ogiya_2021} consider a time-varying spherical potential well to isolate the smooth growth of the host halo and find that the radial action of subhalo orbits decreases by $\sim 10$ per cent over the first few orbits after being accreted.
Both the smooth and merger-induced halo growth are naturally tracked by cosmological DM-only simulations, as those used in our work  ({\sc mdpl2} or {\sc smdpl}) and those considered by \cite{Pullen_SubhaloNonlinearEvolution_2014} and \cite{Yang_SubHaloMCMC_2020}. 
Differences with the latter work may arise because of the lack of massive substructures in their simulations that prevents them from providing strong constraint on the DF model, which affects mainly massive subhaloes.
The halo mass growth is not taken into account in the idealized models adopted by \cite{Velazquez_White_DynFriction_1999} to describe the merging of satellites with a disc galaxy, thus being an additional source of discrepancy between the values of the Coulomb logarithm found.
It is worth noticing that the deepening of the host halo potential can also be produced through adiabatic contraction of the host DM halo as the result of condensation of baryons produced by gas cooling in its centre (e.g., \citealt{BlumenthalFaber_1986}, \citealt{Gnedin_2004}). However, this effect can only be taken into account self-consistently in hydrodynamical simulations, or can be modelled as proposed, for instance, by \cite{Gnedin_2004}. This effect is neither considered in our orbital evolution model nor taken into account by the aforementioned works that give estimates of the Coulomb logarithm, which are based on DM-only simulations. 
Thus, implementing a modelling of adiabatic contraction would not help to alleviate the mismatch in the Coulomb logarithm we have pointed out.

The parameter $ f $, involved in the proximity criterion for mergers, determines the minimum distance an USH can approach the centre of its host before merging with it.
This value gives us an estimate of the size of the central galaxy that will inhabit the DM halo as provided by a SAM.
For the best-fitting parameter $ f = 0.036 $, the assumed size for the central galaxy would be $R_{\mathrm{g}} \sim 0.036 R_{\mathrm{vir}}$. This relation is compatible with the size-virial radius relation found by \cite{Kravtsov_SizeRadiusRelation_2013}, which links the half mass radius of a galaxy with the virial radius of its host halo according to $ r_{1/2} \sim 0.02-0.03~R_{200} $.

As we have seen, the parameter $ \alpha $ controls the efficiency of TS. \cite{Zentner_Bullock_HaloSubstructure_2003} assumes a fixed value $ \alpha = 1 $, while other authors assume TS to be an instantaneous process which implies $\alpha \rightarrow \infty$ \citep{Peniarrubia_Benson_SubhaloDynamicalEvolution_2005}. The best-fitting value $\alpha=3$ we have found in the parameter exploration is consistent with the values obtained by other authors who adjust $ \alpha $ in order to match the results of simulations   (\citealt{Zentner_SubhaloModel_2005}, $\alpha = 2.5$; \citealt{Yang_SubHaloMCMC_2020}, $\alpha = 2.86$; \citealt{Pullen_SubhaloNonlinearEvolution_2014}, $\alpha = 2.5$; \citealt{Gan_SubhaloDynamicalEvolution_2010} $\alpha = 1 -10 $).
Our results suggest that a value $ \alpha > 5 $ would make the 2PCF to be almost totally suppressed at small separations, disfavouring an instantaneous tidal stripping scenario. We arrive  at this conclusion independently of the method used to obtain the best-fitting parameters (namely, using the SHMF and 2PCF as constraints, and via the study of individual orbits).

We analyse the number fraction of both disrupted and merged USHs, which depends on the values of the free parameters of the model. We consider the population of UHSs of any mass (i.e. without applying the mass cut of $10^{10.4} ~ h^{-1}\,\mathrm{M_{\odot}}$) and evaluate different combinations of the values of the parameters $b$, $f$ and $\alpha$, varying them with respect to the fiducial values  $ (f = 0.05, b = 0.5, \alpha = 2.0) $, as those chosen in Fig.~\ref{results:Fig:parameter_variation}. We find that the number fraction of disrupted USHs tends to increase with decreasing redshift, taking values within the ranges 
$\sim 0 - 0.010 $ at $z=2$, $\sim 0 - 0.016 $ at $z=1$, and $\sim 0 - 0.018$ at $z=0$. 
At each redshift, the larger fractions correspond to larger values of $\alpha$ (more efficient TS), as expected;  the maximum value of each range is obtained for $\alpha = 8$. 
The fraction of merged USHs takes values within the ranges 
$\sim 0.002 - 0.047 $ at $z=2$, $\sim 0.001 - 0.031 $ at $z=1$, and $\sim 0.001 - 0.017$ at $z=0$, 
being the maximum value in each range greater for higher redshift.
In general, these 
maximum values are achieved for $b=0.05$, the lowest value tested, which implies a large DF deceleration. This strong deceleration leads to the loss 
of specific angular momentum, 
although a low fraction achieves the minimum value of $0.01\,h^{-1}\,{\rm kpc}\,{\rm km}\,{\rm s}^{-1}$ at each redshift, which represents 
$\sim 10$, $\sim 8$ and $\sim 6$ per cent of the merged USHs at $z = 2, 1$ and $0$, 
respectively. The remaining merged USHs satisfy the proximity merger condition. Therefore, even in this extreme scenario, the proximity merger criterion dominates over the condition of specific angular momentum loss.
The proximity merger condition is fulfilled by 100 per cent of the merged UHSs in most of the remaining parameter combinations. Only when the proximity merger condition becomes too restrictive, i.e for $f=0.01$ (the lowest value tested), this percentage is reduced to $\sim 97$ per cent. Hence, unless the parameter $b$ is sufficiently small (strong DF), all USHs merge because they come close enough to the centre of the host halo. Clearly, for the best-fitting parameters, all USHs merge because they satisfy the proximity merger condition.

We remark that we have not included a tidal heating term in our orbital evolution model. Tidal heating \citep{Spitzer_TidalHeating_1958,Gnedin_TidalHeating_1999,Banik_VdBosch_TidalHeating_2021} is a mechanism produced by rapid changes in tidal forces when a subhalo passes through the pericentre of its orbit. These tidal shocks transfer energy to the satellite, then the subhalo expands and therefore a greater amount of matter is susceptible to be removed via TS.
The impact of this process on the orbital evolution of USHs will be evaluated in a future work.

We recall that the aim of the model presented here is to predict the orbit of an USH considering the main 
dynamical processes that affect it,  but neglecting baryonic effects. Our approach does not intend to improve the realism of DM-only simulations but to reproduce the statistics (SHMF, 2PCF) of high-resolution DM-only simulations from one of lower resolution.
This 
model is designed to be applied on a DM-only cosmological simulation as a pre-processing step before applying a SAM.
With our orbital evolution model, the evolution of satellite galaxies residing in both resolved and unresolved DM subhaloes are treated in a similar way.
Despite the improvements in the behaviour of SHMF and 2PCF shown in this paper (with respect to lower resolution DM-only simulations), our model  (and consequently, its subsequent application to a SAM)
will still suffer from some limitations when compared to hydrodynamical simulations that incorporate the effects of baryons in a self-consistent manner.
The impact of baryons on the dark matter distribution has been studied extensively in the literature \citep[see e.g.][]{Schewtschenko2011_Hydro_DMonly_comparison,Schaller2015,Hellwing2016,Henden2018,Chisari2019,Debackere2020}.
We discuss the aforementioned limitations below.

On the one hand, in SAMs, 
the initial baryonic content of a (sub)halo is assumed as a fraction (equal to the primordial abundance of baryons) of the total (sub)halo mass. However, hydrodynamical simulations have shown that feedback processes can reduce the total (sub)halo mass \citep[see e.g.][]{Cui2014,Sawala2013,Velliscig2014,Castro2021,Eckert2021}. These baryonic effects, which are not taken into account 
by our orbital evolution model of USHs, affect the HMF and SHMF, mainly at low masses. On the other hand, changes in the mass of the objects by baryonic physics also affect the 2PCF.
Higher mass haloes are more strongly clustered, but their clustering signal is affected when feedback processes reduce the stellar content and, consequently, the total mass of the haloes.
Moreover, as generally predicted by hydrodynamical simulations,
the mass density profile of DM (sub)haloes can also be affected by baryonic processes. While the condensation of baryons
\citep[adiabatic contraction;][]{Gnedin_2004}
generates higher densities at small radii, ejection of baryonic matter produced by feedback gives rise to lower densities at large radii.
The effects of TS will be modified in the case of more concentrated distributions of DM in both haloes and subhaloes which will impact in the mass  evolution of the subhaloes. In turn, this will be reflected in the action of dynamical friction leading to a change in the orbit of subhaloes and their spatial distribution.

The effects of baryons on the clustering of subhaloes have been clearly shown in the literature (e.g., \citealt{Weinberg_2008, vanDaalen_2014}) by comparing the 2PCF of subhaloes obtained from hydrodynamical simulations with that given by a DM-only simulation. 
In particular, \cite{vanDaalen_2014} find that subhaloes affected by the presence of baryons present a different clustering at small and intermediate scales. Specifically, for low-mass subhaloes, the combination of cooling and feedback leads to an excess of clustering on very small scales (of the order of the virial radius), and to a reduction of clustering at intermediate scales.
On the other hand, on large scales  
($ r \gg R_{\rm vir} $)
baryons have 
negligible effect 
on the clustering of subhaloes.
One may try to include baryonic effects in a later stage, via post-processing of the DM-only simulation, or with semi-analytic recipes. For example,
the effect of adiabatic contraction could be implemented in our orbital evolution model using, for instance, the analytic fitting functions that describe the transformation of the DM profile presented by \cite{Gnedin_2004}. Such an implementation would require to calibrate the model on the correlation function predicted by hydrodynamical simulations instead of DM-only simulations as considered in our work. However, the purpose of our orbital evolution model is to track USHs mimicking the dynamics of resolved subhaloes in a DM-only simulation for further use by a SAM. Within a SAM, both USHs and resolved subhaloes will be populated by galaxies. If  the effects of baryons on the orbits of USHs were included, the dynamics of orphan satellites (those within USHs with orbits predicted by our model) would be inconsistent with that of satellites residing within resolved subhaloes extracted from a DM-only simulation, which 
takes into account neither the response of dark matter to the condensation of baryons produced by gas cooling in its centre 
nor the ejection of baryonic matter out of the halo due to feedback processes
\citep[but see][]{Beltz-MohrmannBerlind2021}.
Thus, galaxy formation models based on the hybrid technique that combines a DM-only simulation with a SAM has the drawback of producing differences in galaxy clustering, mainly at small scales, with respect to those obtained from hydrodynamical simulations; such differences must be taken into account when comparing the clustering of a galaxy population generated by a SAM with observations.

\section{Conclusions}\label{conclusions}

In this paper, we present a model to track the orbital evolution of unresolved dark matter subhaloes (USHs).
The model includes tidal stripping (TS) effects and dynamical friction (DF). It also takes into account the possible merger of USHs with their host evaluating both its proximity to the centre of the host and the 
reduction of its specific angular momentum to values below $0.01\,h^{-1}\,{\rm kpc}\,{\rm km}\,{\rm s}^{-1}$ 
(i.e. an USH is considered to be merged with its host whichever of these criteria is fulfilled first). The model is characterised by three free parameters: $ b, f, \alpha $, which are involved in the DF deceleration, the proximity merger criterion and the mass removal by TS, respectively.
This model takes into account the main processes that affect the orbital evolution of a 
subhalo, 
and offers a more consistent framework to calculate the dynamical evolution of orphan galaxies in semi-analytic models (SAMs) than other simplified approaches (i.e. following the most-bound particle of their last identified subhalo, or assuming a circular orbit with a decaying radial distance determined from DF).

In order to calibrate the free parameters of the model, 
we consider statistical properties of the population of dark matter (DM) subhaloes obtained from 
{cosmological \em N}-body simulations with different mass resolution, namely, the subhalo mass function (SHMF) and the two-point correlation function (2PCF), and evaluate the convergence between the behaviour of these functions  when tracking the orbital evolution of USHs in the simulation with lower mass resolution. We consider two simulations  of the MultiDark cosmological simulations suite: {\sc smdpl} with box size of $400~h^{-1}\,\mathrm{Mpc}$ on a side and a DM particle mass of $ 9.6 \times 10^7 ~ h^{-1}\,\mathrm{M_{\odot}} $, and  {\sc mdpl2} characterized by a box size of $1~h^{-1}\,\mathrm{Gpc}$ on a side and a DM particle mass of $1.5 \times 10^9~h^{-1}\,\mathrm{M_{\odot}}$.
The proposed calibration procedure calls for the selection of 
boxes 
that are relatively small in volume but representative of the characteristics of the full simulations. 
We consider the following 
subvolumes: 
\textsc{mdpl2} $50~h^{-1}\,\mathrm{Mpc}$ (MD50), \textsc{mdpl2} $100~h^{-1}\,\mathrm{Mpc}$ (MD100) and \textsc{smdpl} $50~h^{-1}\,\mathrm{Mpc}$ (SM50).
The former 
subvolume 
is used to tune the values of the free parameters (calibration box), while the other two are considered in convergence tests.

We adopt the SHMF as a constraint guided by other studies on the orbital evolution of 
subhaloes 
present in the literature (e.g. \citealt{Pullen_SubhaloNonlinearEvolution_2014},  \citealt {Yang_SubHaloMCMC_2020}).
We find that the SHMF $ \tilde{\phi} $ is sensitive to variations in the efficiency of TS (parameter $ \alpha $; see upper-right panel of Figure \ref{results:Fig:parameter_variation}), but it fails to put tight constraints on the parameter $b$ involved in the DF model, and on the parameter $f$ considered in the proximity criterion for mergers, which depend on the position of the USHs (see upper-left and upper-central panels of Figure \ref{results:Fig:parameter_variation}, respectively.)

In this work, we introduce the 2PCF $\xi$ as another constraint to find the best-fitting values of the free parameters of our orbital evolution model.
The 2PCF includes information about the spatial distribution of subhaloes around their hosts, making the calibration procedure more efficient in constraining the values of the free parameters involved in the DF model and merging criterion (see lower-left and lower-central panels of Figure \ref{results:Fig:parameter_variation}).
The clustering information also contributes to constrain the parameter $\alpha$ because the change of mass of the USHs as a result of mass removal by TS have direct impact on their merging time-scales, i.e. a lower mass implies a lower dynamical friction force (equation \ref{model:Eq:Chandrasekhar1943}). Hence, in contrast to the SHMF, 
{\em the 2PCF plays a key role in constraining the three free parameters of our orbital evolution model.}
The exploration of the parameter space of the model gives the following best-fitting set: $ b= 0.3, f= 0.036, \alpha= 3.0$.

To assess the robustness of these best-fitting parameters, we have performed a comparison between the orbits followed by a set of resolved subhaloes in the SM50 box that are reliable according to the criteria presented in \cite{vdBosch_Ogiya_2018}, and the orbits of those same resolved subhaloes predicted by our orbital evolution model. We have defined cost functions that quantify the deviations between the predicted orbit and the real one (i.e., the one followed by the subhalo in SM50), and  estimate the parameters $\alpha$ and $b$ that best describe the real orbit (i.e., the ones that minimize the cost functions). We carry out the analysis for different circularity bins, selecting only those subhaloes whose circularities are stable (i.e. the circularity does not vary significantly along the orbit). 
The values of the parameters found from the analysis of the real orbits of reliable subhaloes provided by the cosmological simulation are consistent with those found by applying the approach presented here (using SHMF and 2PCF as constraints), suggesting that our procedure is appropriate to tune the free parameters of the orbital evolution model used to describe the orbits of USHs.

From the convergence test carried out by tracking the orbits of USHs present in the boxes MD100 and SM50, we evaluate the goodness of the selected calibration box MD50, and the performance of our orbital evolution model characterised by the best-fitting parameters (see Figs.~\ref{results:Fig:BestFit_HMF} and~\ref{results:Fig:BestFit_2PCF}).
On one hand, the similarity of both the SHMF and the 2PCF obtained for the boxes MD50 and MD100 with and without USHs supports the choice of the size of the calibration box, that is, a 
subvolume 
with side-length of  $50~h^{-1}\,\mathrm{Mpc}$ is adequate enough to carry out the calibration procedure.
On the other hand, the very good agreement achieved between the SHMF and the 2PCF obtained after applying the model to the 
subvolume 
of the lower-resolution simulation ({\sc mdpl2}) and those extracted from the full {\sc smdpl} simulation corroborates the current implementation, both in terms of the physical processes considered and the calibration procedure applied. 
Indeed, the mean fractional differences, taking the results of the full {\sc smdpl} simulation as a reference, is $\sim 10$ per cent for both the SHMF and the 2PCF over the mass and scale ranges analysed.

A very interesting point that emerges from this convergence test is that the mass resolution of the {\sc smdpl} simulation seems to be high enough to obtain the right behaviour of both the SHMF and the 2PCF for the sample of subhaloes considered, i.e with masses greater than the mass cut adopted ($10^{10.4} ~ h^{-1}\,\mathrm{M_{\odot}}$). This can be inferred from the small fractional differences between the results obtained for the SM50 box with and without the inclusion of USHs, which are only noticeable at the smallest masses (in the SHMF) and smallest scales (in the 2PCF). 
Thus, DM-only cosmological simulations with mass resolution similar to that of the {\sc smdpl} simulation, that is $\sim 10^8 ~ h^{-1}\,\mathrm{M_{\odot}}$, would not require the inclusion of USHs to have a population of subhaloes characterized by statistical functions with the right behaviour, at least for the SHMF and 2PCF and the ranges of masses and scales considered in this work. The need for following the orbital evolution of those few USHs would become evident when evaluating statistics of the galaxy population as modelled by a SAM, since the merger history of orphan galaxies inhabiting those USHs would be crucial in defining the different properties of the baryonic components of galaxies. This points to the question about the importance of including the evolution of orphan galaxies in SAMs, a certainly important issue that deserves further investigation.

\section*{Acknowledgements}

The authors thank the referee and the Scientific Editor Prof. Joop Schaye for useful comments and suggestions that help to greatly improve this manuscript.

FMD and CGS are supported  by the National Agency for the Promotion of Science and Technology (ANPCYT) of Argentina grant PICT-2016-0081; and grants G140, G157 and G175 from UNLP.
SAC and IDG acknowledges funding from {\it Consejo Nacional de Investigaciones Cient\'{\i}ficas y T\'ecnicas} (CONICET, PIP-0387), {\it Agencia Nacional de Promoci\'on de la Investigaci\'on, el Desarrollo Tecnol\'ogico y la Innovaci\'on} (Agencia I+D+i, PICT-2018-03743), and {\it Universidad Nacional de La Plata} (G11-150), Argentina. CVM acknowledges support from ANID/FONDECYT through grant 3200918, and support from the Max Planck Society through a Partner Group grant.

The CosmoSim database used in this paper is a service by the Leibniz-Institute for Astrophysics Potsdam (AIP).
The MultiDark database was developed in cooperation with the Spanish MultiDark Consolider Project CSD2009-00064.

\section*{Data Availability}

The data underlying this article will be shared on reasonable request to the corresponding author. The simulations used in this paper are publicly available at  \url{https://www.cosmosim.org/}.



\bibliographystyle{mnras}
\bibliography{bibliography} 



\appendix

\section{Ingredients of the orbital evolution model for a NFW density profile}\label{appendixA}

In numerical studies, the detected dark matter haloes are well described by a double power-law function of the radius, first introduced by \cite{Navaro_NFW_1997}. Here we describe the Navarro-Frenk-White (NFW) density profile, and we give expressions for the dynamical friction formula and tidal radius equation corresponding to this particular profile.

The NFW density profile is given by the following expression

\begin{equation}
    \frac{\rho(r)}{\rho_{\mathrm{c}}}  = \frac{\delta_{\mathrm{char}}}{\frac{r}{r_{\mathrm{s}}} \left( 1 + \frac{r}{ r_{\mathrm{s}} } \right)^2 } ~.
    \label{appendixA:Eq:NFW_profile}
\end{equation}
\noindent The parameters of this expression are the scale radius $ r_{\mathrm{s}} $ and the characteristic density $ \delta_{\mathrm{char}} $. The concentration parameter is defined as

\begin{equation}
    c = \frac{R_{\mathrm{vir}}}{r_{\mathrm{s}}} ~,
    \label{appendixA:Eq:concentration}
\end{equation}
\noindent where $ R_{\mathrm{vir}} $ is the virial radius of the halo, defined as the distance from the center of the halo within which the mean density is $ \Delta $ times the critical density $ \rho_{\mathrm{c}} $. The value of the virial overdensity is often assumed to be $ 178 $, predicted by the spherical collapse model. However, numerical simulations typically use $ \Delta = 200 $.

We also introduce the dimensionless distance $ s $ and the function $ g(c) $, which often appears in calculations involving the NFW profile,

\begin{equation}
    s = \frac{r}{R_{\mathrm{vir}}} ~,
    \label{appendixA:Eq:s}
\end{equation}
\begin{equation}
    g(c) = \frac{1}{\ln(1 + c) - c/(1 + c)} ~.
    \label{appendixA:Eq:gc}
\end{equation}
With the above definitions, the density profile becomes

\begin{equation}
     \frac{\rho(s)}{\rho_{\mathrm{c}}} = \frac{ c^{2} \, g(c) \, \Delta}{ 3s \, (1 + c s)^2 } ~.
     \label{appendixA:Eq:NFW_dimensionless}
\end{equation}
The mass of a DM halo is usually defined as the mass within the virial radius

\begin{equation}
     M_{\mathrm{vir}} = 4/3 \pi \, R_{\mathrm{vir}}^{3} \, \Delta \, \rho_{c} ~.
     \label{appendixA:Eq:NFW_Mvir}
\end{equation}
Then the mass profile in units of the virial mass is

\begin{equation}
    \frac{M(s)}{M_{\mathrm{vir}}} = g(c) \left[ \ln(1 + c s) - \frac{cs}{1 + c s} \right] ~.
    \label{appendixA:Eq:NFW_Ms}
\end{equation}
Using the same definitions, we can express the gravitational potential as

\begin{equation}
    \frac{\Phi(s)}{V_{\mathrm{c}}^2} = - g(c) \frac{\ln(1 + c s)}{s} ~,
    \label{appendixA:Eq:NFW_potential}
\end{equation}
\noindent where 

\begin{equation}
    V_{\mathrm{c}} = \sqrt{G M_{\mathrm{vir}}/R_{\mathrm{vir} } }
    \label{appendixA:Eq:NFW_Vcircl}
\end{equation}

\noindent is the circular velocity at $ r = R_{\mathrm{vir}}$.

The orbital model presented in this work (Section \ref{model}) requires the computation of the acceleration and tidal radius of 
USHs 
orbiting within a host halo. 
In order to simplify the computational implementation of the model (Section \ref{subsection:OrbitModelSummary}), 
for an USH of mass $M_{\rm ush}$, located at a distance $r_{\rm ush}$ of its host and moving with relative velocity $V_{\rm ush}$, we introduce the following dimensionless quantities

\begin{equation}
    m_{\rm ush} = \frac{M_{\rm ush}}{ M_{\rm host}}~,
    \label{appendixA:Eq:m_dimless}
\end{equation}
\begin{equation}
    s_{\rm ush} = \frac{r_{\rm ush}}{R_{\rm host}}~,
    \label{appendixA:Eq:s_dimless}
\end{equation}
\begin{equation}
    v_{\rm ush} = \frac{V_{\rm ush}}{V_{\rm c,host}}~,
    \label{appendixA:Eq:v_dimless}
\end{equation}
\begin{equation}
    \omega_{\rm ush} = \frac{\Omega_{\rm ush}}{V_{\rm c,host}/R_{\rm host}}~,
    \label{appendixA:Eq:omega_dimless}
\end{equation}

where $ M_{\rm host} $, $ R_{\rm host} $ and $ V_{\rm c,host} $ 
are the mass, radius and circular velocity of the host, respectively.

The radial acceleration of a subhalo at a distance $ r $ is given by $ a(r) = - G M(<r)/r^2 $, 
using dimensionless quantities
(\ref{appendixA:Eq:m_dimless}), 
(\ref{appendixA:Eq:s_dimless}),  (\ref{appendixA:Eq:v_dimless}), 
and assuming a NFW mass profile (equation \ref{appendixA:Eq:NFW_Ms}) with concentration $ c $ for the host, we can express the $i$-component of the acceleration in Cartesian coordinates as

\begin{equation}
    a_{{\rm ush}, i} = - \frac{ V_{\rm c,host}^2}{R_{\rm host}} 
    g(c) \left[ \frac{\ln(1 + c \, s_{\rm ush})}{s_{\rm ush}^2} - \frac{c}{(1 + c \, s_{\rm ush}) \, s_{\rm ush}}  \right] \frac{s_{{\rm ush}, i}}{s_{\rm ush}} ~. \label{appendixA:Eq:NFW_acceleration}
\end{equation}

Analogously, the dynamical friction acceleration which is given by the Chandrasekhar formula (equation \ref{model:Eq:Chandrasekhar1943}), can be expressed in the following form

\begin{equation}
\begin{split}
    a_{{\rm ush},i}^{\rm df} = - &\frac{ V_{\rm c,host}^2}{R_{\rm host}} 
    \frac{ m_{\rm ush} \, \ln{\Lambda} }{v_{\rm ush}^2} \frac{c^2 g(c)}{s_{\rm ush} (1 + c \, s_{\rm ush})^2} ~\times \\
    &\left[ \mathrm{erf}(X) - \frac{2 X}{\sqrt{\pi}} \mathrm{exp}{(- X^2)} \right] \frac{v_{{\rm ush}, i}}{v_{\rm ush}} ~.
    \label{appendixA:Eq:NFW_DF}
\end{split}
\end{equation}

\noindent Here 
$ X = V_{\rm ush} / (\sqrt{2} \sigma)$ 
with $ \sigma $ the velocity dispersion of dark matter particles in the host halo.

Finally, for estimating the mass to be removed via TS we need to compute the tidal radius, i.e.

\begin{equation}
    r_{\mathrm{t}} = \left( \frac{G M_{\rm ush}}{ \Omega_{\rm ush}^{2} - d^2 \Phi / d r^2 } \right)^{1/3} ~.
    \label{appendixA:Eq:NFW_King}
\end{equation}

\noindent Taking into account the extended mass profile of the host halo, then for $ d^2 \Phi / d r^2$ we have

\begin{equation}
    \frac{d^2 \Phi}{d r^2} 
    \biggr\rvert_{r = r_{\mathrm{ush}}} = - \frac{2 G M_{\mathrm{host}}(< r_{\mathrm{ush}})}{r_{\mathrm{ush}}^3} + 4 \pi G \rho_{\rm host}(r_\mathrm{ush}) ~,
    \label{appendixA:Eq:NFW_dPhi}
\end{equation}

\noindent where 
$r_{\rm ush}$ 
is the position of the satellite within the host halo and 
$M_{\mathrm{host}}(< r_{\rm ush})$ 
is the mass of the host contained within the distance determined by the satellite position \citep{Pullen_SubhaloNonlinearEvolution_2014}.
Using the definitions (\ref{appendixA:Eq:m_dimless}), 
(\ref{appendixA:Eq:s_dimless}),  (\ref{appendixA:Eq:v_dimless}), 
(\ref{appendixA:Eq:omega_dimless}) and the expresions 
(\ref{appendixA:Eq:NFW_dimensionless}) and (\ref{appendixA:Eq:NFW_Ms}) for $\rho_{\mathrm{host}}(s) $ and $M_{\mathrm{host}}(s) $, respectively,
we obtain the tidal radius

\begin{equation}
    r_{\mathrm{t}} = R_{\rm host} 
    \left(
    \frac{m_{\rm ush} \, s_{\rm ush}^3 }{ 
    \omega_{\rm ush}^2 \, s_{\rm ush}^3 + 2 \, g(c) \left[\ln{(1 + c \, s_{\rm ush})} -\frac{c \,  s_{\rm ush}}{1 + c \, s_{\rm ush}} \right] - \frac{c^2 \, g(c) \, s_{\rm ush}^2}{(1 + c \, s_{\rm ush})^2}}
    \right)^{1/3} ~.
    \label{appendixA:Eq:NFW_rt}
\end{equation}


\bsp 
\label{lastpage}

\end{document}